\documentclass[twocolumn]{aastex61}

\shorttitle{\textit{Herschel}/PACS OH Spectroscopy}
\shortauthors{Runco et al.}

\begin{document}

\title{\textit{Herschel}/PACS OH Spectroscopy of Seyfert, LINER, and Starburst Galaxies\footnote{\textit{Herschel} is an ESA space observatory with science 
\newline instruments provided by European-led Principal Investigator 
\newline consortia and with important participation from NASA.}}

\correspondingauthor{Jordan N. Runco}
\email{jrunco@astro.ucla.edu}

\author[0000-0003-4852-8958]{Jordan N. Runco}
\affil{Physics \& Astronomy Department, University of California: Los Angeles, 430 Portola Plaza, Los Angeles, CA 90095, USA}

\author[0000-0001-6919-1237]{Matthew A. Malkan}
\affil{Physics \& Astronomy Department, University of California: Los Angeles, 430 Portola Plaza, Los Angeles, CA 90095, USA}

\author[0000-0001-9490-899X]{Juan Antonio Fern\'andez-Ontiveros}
\affiliation{Istituto di Astrofisica e Planetologia Spaziali, INAF-IAPS, Via Fosso del Cavaliere 100, I-00133 Roma, Italy}

\author[0000-0001-8840-1551]{Luigi Spinoglio}
\affiliation{Istituto di Astrofisica e Planetologia Spaziali, INAF-IAPS, Via Fosso del Cavaliere 100, I-00133 Roma, Italy}

\author[0000-0002-4005-9619]{Miguel Pereira-Santaella}
\affiliation{Department of Physics, University of Oxford, Keble Road, Oxford, OX1 3RH, UK}
\affiliation{Centro de Astrobiolog\'ia (CSIC-INTA), Ctra. de Ajalvir, Km 4, 28850, Torrej\'on de Ardoz, Madrid, Spain}

\begin{abstract}
We investigated the 65$\mu$m, 71$\mu$m, 79$\mu$m, 84$\mu$m, 119$\mu$m, and 163$\mu$m OH doublets of 178 local (0 \textless \space $z$ \textless \space 0.35) galaxies. They were observed using the $Herschel$/PACS spectrometer, including Seyfert galaxies, LINERs, and star-forming galaxies. We observe these doublets exclusively in absorption (OH71), primarily in absorption (OH65, OH84), mostly in emission (OH79), only in emission (OH163) and an approximately even mix of the both (OH119).  In 19 galaxies we find P-Cygni or reverse P-Cygni line profiles in the OH doublets. We use several galaxy observables to probe spectral classification, brightness of a central AGN/starburst component, and radiation field strength.  We find that OH79, OH119, and OH163 are more likely to display strong emission for bright, unobscured AGN.  For less luminous, obscured AGN and non-active galaxies, we find populations of strong absorption (OH119), weaker emission (OH163), and a mix of weak emission and weak absorption (OH79).  For OH65, OH71 and OH84, we do not find significant correlations with the observables listed above.  We do find relationships between OH79 and OH119 with both the 9.7$\mu$m silicate feature and Balmer decrement dust extinction tracers in which more dust leads to weaker emission / stronger absorption.  The origin of emission for the observed OH doublets, whether from collisional excitation, or from radiative pumping by infrared photons, is discussed.
\end{abstract}


\section{Introduction} \label{sec:intro}

The different modes of feedback from active galactic nuclei (AGNs) are thought to play a vital role in the evolution of massive galaxies \citep{som15}, by quenching star formation and the growth of supermassive black holes in their host galaxy \citep{di05}.  

Molecular gas is a direct probe of AGN driven outflows \citep{spo13}, which could be capable of displacing a large fraction of the molecular interstellar medium of the host galaxy \citep{stu11, vei13}.  Among the different molecules that have been used to trace outflows (CO in \citealt{fer10}; OH in \citealt{fis10, stu11}, HCN, HCO$^{+}$, and HNC in \citealt{aal12}), OH is considered one of the best diagnostic tools due to its large dipole moment (1.67 Debye) and thus its fast radiative rates.  The critical densities of the OH transitions are high ($\sim$10$^{9}$cm$^{-3}$), which makes them sensitive to radiative pumping \citep{spo13}.  There are 14 OH transitions between 34$\mu$m and 163$\mu$m that arise from the eight lowest energy rotational levels.  Those relevant for this study are shown on the OH Grotrian diagram in Figure \ref{fig:energy_diagram}.  The number of infrared (IR) lines available is another value of OH, as it can probe a broad set of conditions in the molecular gas \citep{spo13}.

The OH doublets can vary widely in their properties.  Galaxies have been observed with these OH doublets primarily in emission \citep{spi05}, primarily in absorption \citep{keg19, gon04}, a combination of the two \citep{bra99, gon08}, and even P-Cygni features \citep{fis10}.  P-Cygni profiles indicate unambiguously the presence of molecular gas outflows.

Dedicated radiative transfer modeling including several OH and H$_{2}$O transitions have been performed for a few individual galaxies (e.g. Arp 220 and Mrk 231; \citealt{gon04, gon08}).  These studies have utilized data from the \textit{Infrared Space Observatory} Long Wave Spectrometer (\textit{ISO}/LWS; \citealt{cle96, kes96}) and the Photoconductor Array Camera and Spectrometer (PACS; \citealt{pog10}) instrument on the \textit{Herschel Space Observatory} \citep{pil10}.  The capabilities of the \textit{Herschel}/PACS instrument have led to many studies using the OH line profiles to look for molecular outflow signatures (e.g. \citealt{fal15, fis10, gon12, gon14, gon15, gon17a, sto16, stu11, vei13}).

\begin{deluxetable*}{ccccc}
\tablecolumns{5}
\tablecaption{Fine Structure Transitions \label{tab:oh_lines}}
\tablehead{
  \colhead{Line} &
  \colhead{Energy Levels} &
  \colhead{Wavelength} & 
  \colhead{$\Delta E_{ul}/k$} &
  \colhead{$A_{ul}$} \\
  \colhead{ } & \colhead{} & \colhead{($\mu$m)} & \colhead{(K)} & \colhead{(s$^{-1}$)} \\
  \colhead{(1)} & \colhead{(2)} & \colhead{(3)} & \colhead{(4)} & \colhead{(5)}}
\startdata
OH65& $^2 \Pi_{3/2}$(J = 9/2) $\longleftrightarrow$ $^2 \Pi_{3/2}$(J = 7/2) & 65.132, 65.279 & 512.1, 510.9 & 1.276, 1.267 \\
OH71& $^2 \Pi_{1/2}$(J = 7/2) $\longleftrightarrow$ $^2 \Pi_{1/2}$(J = 5/2) & 71.171, 71.215 & 617.6, 617.9 & 1.014, 1.012 \\
OH79& $^2 \Pi_{1/2}$(J = 1/2) $\longleftrightarrow$ $^2 \Pi_{3/2}$(J = 3/2) & 79.116, 79.179 & 181.9, 181.7 & 0.03606, 0.03598 \\
OH84& $^2 \Pi_{3/2}$(J = 7/2) $\longleftrightarrow$ $^2 \Pi_{3/2}$(J = 5/2) & 84.420, 84.596 & 291.2, 290.5 & 0.5235, 0.5202 \\
OH119& $^2 \Pi_{3/2}$(J = 5/2) $\longleftrightarrow$ $^2 \Pi_{3/2}$(J = 3/2) & 119.234, 119.441 & 120.7, 120.5 & 0.1388, 0.1380 \\
OH163& $^2 \Pi_{1/2}$(J = 3/2) $\longleftrightarrow$ $^2 \Pi_{1/2}$(J = 1/2) & 163.015, 163.396 & 270.2, 269.8 & 0.06483, 0.06450 \\
\enddata
\tablecomments{Col. (1): Name of OH doublet.  Col. (2): Energy transition for each doublet.  Col. (3): Wavelength for both transitions in the doublet. Col. (4): Energy of the upper level for each transition.  Col. (5): Einstein A coefficient for each transition.  This data is from the LAMBDA database \citep{sch05}.}
\end{deluxetable*}

Some relationships between far-IR OH molecular line properties and host galaxy properties have already been found.  Many of these studies have primarily focused on the 119$\mu$m OH doublet.  This transition has been found to display stronger emission in more luminous AGNs \citep{vei13, sto16}.  More luminous AGNs also tend to have higher terminal outflow velocities for the 79$\mu$m \citep{stu11} and 119$\mu$m \citep{vei13} OH doublets.  Dust has been shown to influence the 119$\mu$m OH doublet as well.  Using the 9.7$\mu$m silicate feature as a tracer for the dust content of the galaxy, \citet{vei13} and \citet{sto16} show that stronger 119$\mu$m OH doublet emission is correlated with the 9.7$\mu$m silicate feature in either weak absorption or emission, which indicates less dust.  A dusty galaxy corresponds to a weak emission or absorption feature for the 119$\mu$m OH doublet.  The 65$\mu$m OH doublet is also shown to show stronger absorption with increasing 9.7$\mu$m silicate absorption \citep{gon15}.  

The main objective of this paper is to present a large sample of 178 local galaxies, and quantify the emission or absorption of six OH doublets in the far-IR range $-$ 65$\mu$m, 71$\mu$m, 79$\mu$m, 84$\mu$m, 119$\mu$m, and 163$\mu$m $-$ by measuring the equivalent width (EW) of the transitions.  We chose to use the EW to quantify the transitions in the sample because the EW measures the relative strength of the OH doublet relative to the continuum emission.  There are also nine galaxies in the sample with 53.3$\mu$m observations and four with 98.7$\mu$m observations; however, due to the combination of a lack of statistics and a lack of resolution to separate the peaks, in particular for the 99$\mu$m doublet, we do not include them in this study.  We will use our large sample to look for statistical trends in the data, and investigate possible correlations between EW(OH) and host galaxy properties similar to the studies discussed above.  However, in our study, we will investigate a more expansive range of galaxy properties and higher number of OH transitions to gain a more complete understanding of OH properties in galaxies.  We also span a wide variety of galaxy spectral types, with the goal of obtaining a more complete understanding of the OH spectra.  Since we still do not have a full theoretical understanding of the observed OH spectra, here we take an empirical approach to uncover which other observed properties of galaxies may predict the conditions in their molecular gas.  

This paper is organized as follows.  Section \ref{sec:observations} describes the sample selection, the observations, the data reduction, and ancillary data collected.  Section \ref{sec:analysis_and_results} describes the data analysis and the quantities derived from the data.  Section \ref{sec:discussion} provides a discussion of the results.  Section \ref{sec:summary} provides a summary of the paper.  Finally, the Appendix provides statistical information for fits discussed in Section \ref{sec:analysis_and_results}.

\section{Observations} \label{sec:observations}

\begin{figure}[ht!]
    \centering
    \includegraphics[scale=0.36]{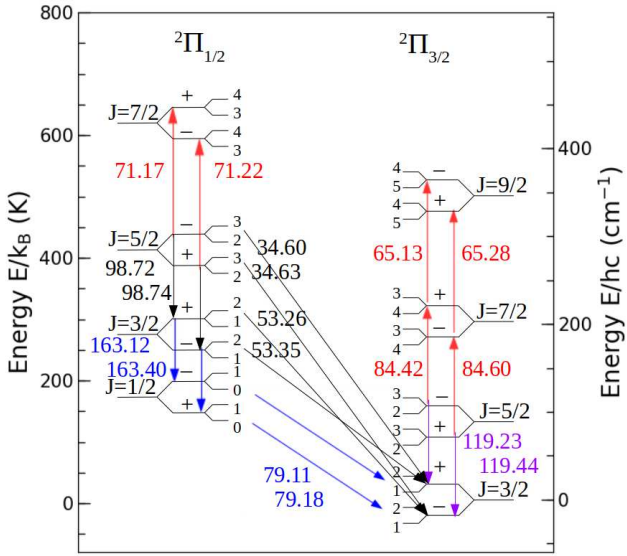}
    \caption{OH rotational Grotrian diagram showing the transitions discussed in this paper.  Each rotational level is split into two sublevels due to $\Lambda$ doubling and each sublevel is split again due to the hyperfine structure. The rotational ladder has two branches due to spin splitting ($\Lambda$ + s, $\Lambda$ $-$ s) where $\Lambda$ = 1 and s = 1/2. The integer next to the hyperfine branches of each energy level is the total angular-momentum quantum number, F.  The red, blue, purple, and black numbers are the wavelength of the transition in units of $\mu$m.  Red indicates that the doublet was seen primarily in absorption, blue indicates primarily emission, purple indicates that neither emission or absorption was strongly favored, and black indicates transitions not analyzed in this study but are mentioned in Section \ref{sec:discussion}.}
    \label{fig:energy_diagram}
\end{figure}

This paper investigates six of the 14 OH transitions in the far-IR range: 65$\mu$m, 71$\mu$m, 79$\mu$m, 84$\mu$m, 119$\mu$m, and 163$\mu$m.  These six are selected because they lie in the wavelength range covered by PACS and have the best chance of being detected.  Hereafter, we refer to these transitions as OH65, OH71, OH79, OH84, OH119, and OH163.  All of these transitions are doublets.  Table \ref{tab:oh_lines} shows details of the transitions for these doublets, and Figure \ref{fig:energy_diagram} provides the OH Grotrian diagram.  In the next section, we describe the OH measurements obtained from Herschel spectra and additional ancillary measurements of those galaxies, which may be correlated with EW(OH).  

\subsection{$Herschel$/PACS} \label{subsec:PACS}

A sample of 178 local ($z$ \textless \space 0.35) galaxies were selected from the \textit{Herschel Science Archive}, which were observed with the PACS instrument \citep{pog10} on the \textit{Herschel Space Observatory} \citep{pil10}.  PACS used an integral field spectrograph with a $5 \times 5$ square spaxel detector, which covers a field of view of $47" \times 47"$.  The spectral resolution is wavelength-dependent, with a range of $R$ = 1000 $-$ 4000 ($\Delta\nu$ = 75 $-$ 300 km s$^{-1}$).  PACS spans a wavelength range of 51 $-$ 206$\mu$m.  

To obtain the largest sample for statistical analysis, we follow a similar selection procedure as that described in \citet{ont16}. In \citet{ont16}, 170 AGN with available mid-IR spectra from \textit{Spitzer}/IRS \citep{hou04, wer04} and far-IR fine structure lines from \textit{Herschel}/PACS were selected from the \textit{Herschel Science Archive}.  A comparison sample of 20 starburst galaxies were also selected.  In this study, the sample of AGN (163 galaxies) and starbursts (15 galaxies) is slightly smaller.  We do not include galaxies from \citet{ont16} in our sample that do not have observations for at least one of the six OH doublets that we analyze in this paper.  However, there are 14 AGN added to this sample that are not in \citet{ont16} because their selection was based on the detection of fine-structure lines, instead of the OH doublets.  Therefore, the samples in this study and \citet{ont16} have significant overlap, with only minor differences based on the specific focus of the paper.

The observations originated from many different Herschel proposals.  Each project had its own criteria for selecting galaxies, and here are a few of the main reasons why galaxies were targeted:
\begin{itemize}
    \item Many observed galaxies are very bright, especially in the IR.  These objects are ultraluminous infrared galaxies (ULIRGs).  The 60$\mu$m flux was often chosen for these flux-limited samples.  This favors dusty galaxies with high obscuration, included obscured AGN.  There were a variety of proposals that targeted these objects (e.g. KPOT\_pvanderw\_1, OT1\_dfarrah\_1, OT2\_jgracia\_1, OT2\_larmus\_2). 
    \item Previously well-studied galaxies were selected.  For example, galaxies were selected from previous surveys (e.g. \textit{Spitzer} Infrared Nearby Galaxy Survey (SINGS; \citealt{ken03}) and Burst Alert Telescope (BAT; \citealt{bar05})).
    \item Galaxies were targeted based on their optical spectroscopic classification.  This includes Seyferts and starburst galaxies (see \citealt{ont16} and references therein).
\end{itemize}

\begin{deluxetable*}{cccccc}
\tabletypesize{\scriptsize}
\tablecolumns{6}
\tablecaption{The Sample \label{tab:sample}}
\tablehead{
  \colhead{Object Name} &
  \colhead{R.A.} & 
  \colhead{Dec.} &
  \colhead{$z$} &
  \colhead{Optical Spectral} &
  \colhead{Transitions Observed} \\
  \colhead{ } & \colhead{(J2000)} & \colhead{(J2000)} & \colhead{ } & \colhead{Classification} & \colhead{($\mu$m)} \\
  \colhead{(1)} & \colhead{(2)} & \colhead{(3)} & \colhead{(4)} & \colhead{(5)} & \colhead{(6)}}
\startdata
Mrk334 & 00h03m09.6038s & +21d57m36.8064s & 0.021945 & Seyfert-1.8 & 79 \\
IRAS00182-7112 & 00h20m34.7210s & $-$70d55m26.2488s & 0.326999 & Seyfert-2 & 79,84,119 \\
IRAS00198-7926 & 00h21m53.6141s & $-$79d10m07.9572s & 0.0728 & Seyfert-2 & 79 \\
NGC253 & 00h47m33.0727s & $-$25d17m18.9960s & 0.000811 & Starburst & 65,71,79,84,119,163 \\
Mrk348 & 00h48m47.1468s & +31d57m25.1280s & 0.015034 & Seyfert-1h & 84 \\
IZw1 & 00h53m34.9236s & +12d41m35.9232s & 0.0612 & Seyfert-1n & 65,79,119,163 \\
IRAS00521-7054 & 00h53m56.2310s & $-$70d38m04.2216s & 0.0689 & Seyfert-1h & 79 \\
ESO541-IG12 & 01h02m17.3818s & $-$19d40m08.6556s & 0.056552 & Seyfert-2 & 79 \\
IRAS01003-2238 & 01h02m49.9894s & $-$22d21m57.2616s & 0.117835 & Seyfert-2 & 79,119 \\
NGC454E & 01h14m24.9300s & $-$55d23m49.2936s & 0.012158 & Seyfert-2 & 79 \\
\enddata
\tablecomments{
  Col. (1): Target name.
  Col. (2): Right ascension (collected from the 2MASS Point Source Catalog).
  Col. (3): Declination (collected from the 2MASS Point Source Catalog).
  Col. (4): Redshifts gathered from the NASA/IPAC Extragalactic Database (NED).
  Col. (5): Optical spectral type taken from the \citet{ver10} Catalog.
  Col. (6): The OH doublets that were observed for each object.}
\tablecomments{Table \ref{tab:sample} is published in its entirety in the machine-readable format.
      A portion is shown here for guidance regarding its form and content.}
\end{deluxetable*}

Because there is a large variety in the sample selection process for choosing galaxies, we believe that our final sample of 178 galaxies does not have a strong bias towards any particular subset of galaxies.  However, there are two exceptions: there are no dwarf galaxies in our sample and galaxies known to be particularly luminous in the far-IR received somewhat more attention.  Therefore, the results found in this paper can be considered reflective of how the entire population of low-redshift, non-dwarf, IR-luminous galaxies behave.

Table \ref{tab:sample} contains the optical spectral type of every galaxy in the sample, which were obtained from the \citet{ver10} catalog.  Note that while some galaxies have multiple AGN classifications in the catalog, we only include the first one listed.  Seyfert types-1, -1.2, -1.5, and -1.8 have both broad H$\alpha$ and broad H$\beta$ components.  To distinguish between them, the \citet{ver10} Catalog used a quantitative method based on the ratio of the H$\beta$ and [O~\textsc{III}]$\lambda$5007 line fluxes introduced by \citet{win92}.  Seyfert-1.9's have a broad H$\alpha$ component visible but not broad H$\beta$, and Seyfert-2's have neither broad components \citep{ost77, ost81}.  Seyfert-1n are narrow lined Seyfert-1's, which have optical emission spectra similar to Seyfert-1's but have narrower Balmer lines \citep{ost85, goo89}.  Seyfert-1h's have broad Balmer lines that are only detected in polarized light, indicating that there is a hidden broad-line-region that is completely obscured \citep{ant85, mil90, tra92}.  Seyfert-1i's have a highly reddened broad-line-region which is revealed by a broad Pa $\beta$ line in the spectra \citep{goo94}.  Low Ionization Nuclear Emission-line Regions (LINERs) have an optical spectrum dominated by low ionization emission lines, which have widths similar to the narrow-line-region of Seyfert galaxies but at a much lower luminosity \citep{hec80}.  If broad Balmer lines are observed, the galaxy is classified as a LINERb \citep{ver10}.

\begin{figure*}[t!]
    \centering
    \includegraphics[scale=0.55]{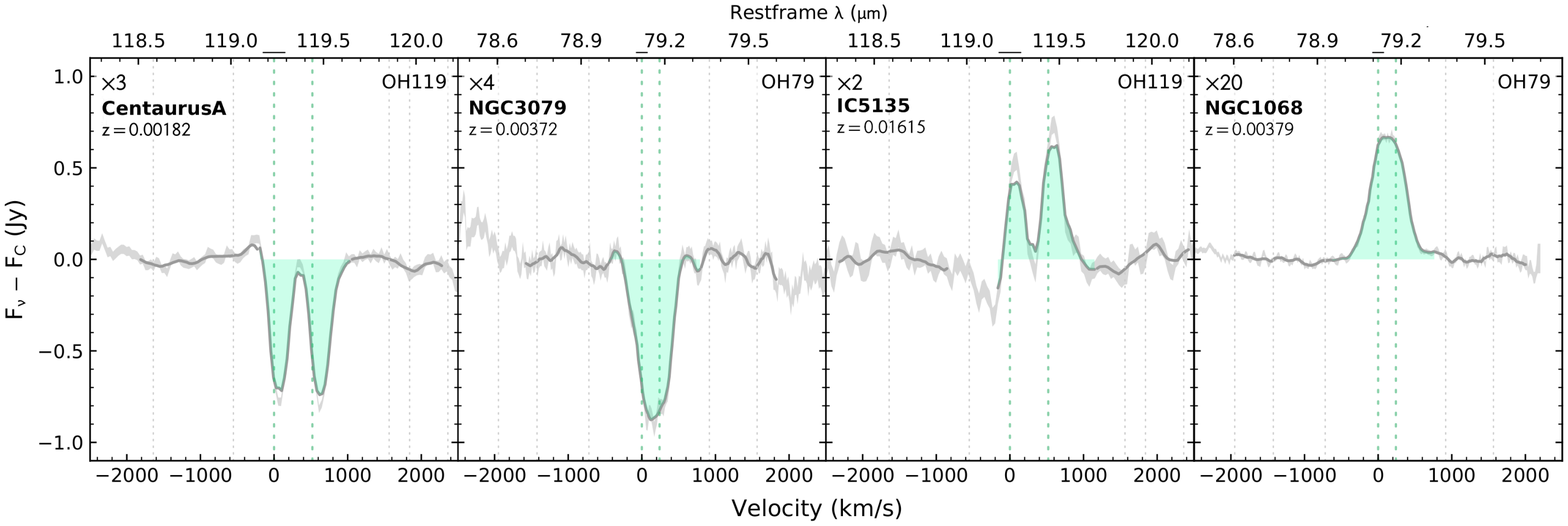}
    \includegraphics[scale=0.55]{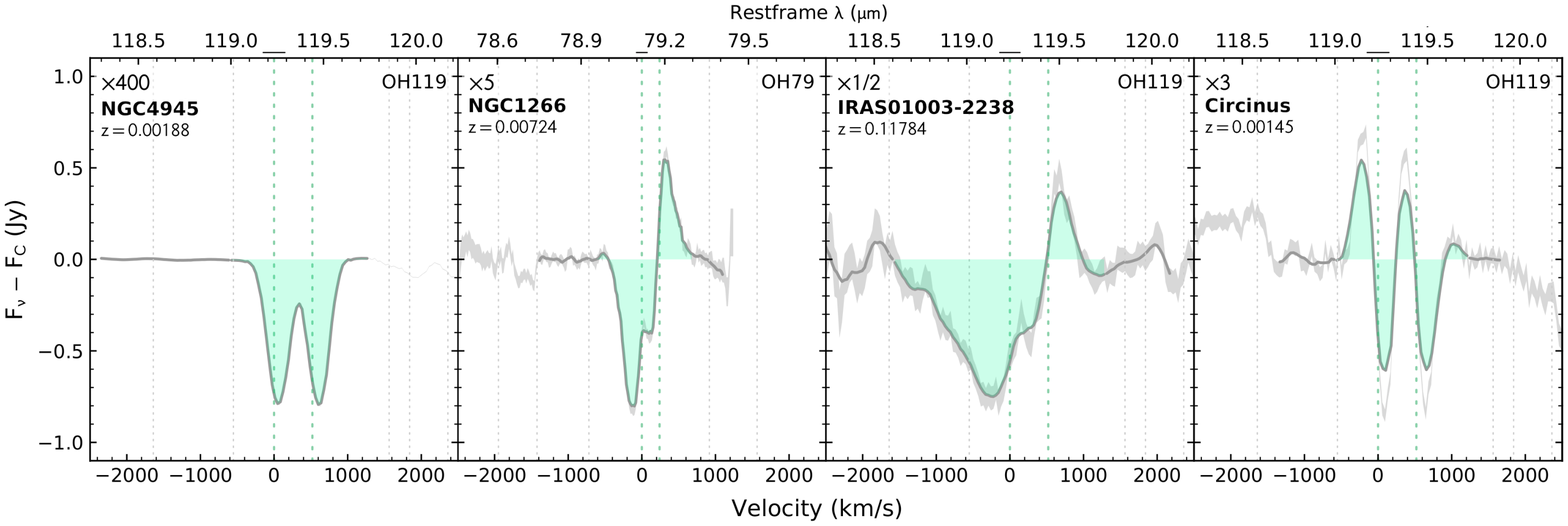}
    \caption{Examples of the variety of spectra observed with PACS.  These spectra are the central spaxel of the observations of the $5 \times 5$ array. From left to right: Top: doublet absorption (Centaurus A: Seyfert-2); absorption where the doublet is blended into one line (NGC 3079: LINER); doublet emission (IC 5135: Seyfert-1.9); emission where the doublet is blended into one line (NGC 1068: Seyfert-1h). Bottom: doublet absorption (NGC 4945: Seyfert-2); P-Cygni profile with the emission and absorption components approximately equal in strength (NGC 1266: LINER); P-Cygni profile with a stronger absorption component than emission (IRAS 01003-2238: Seyfert-2); reverse P-Cygni profile with emission and absorption components approximately equal in strength (Circinus: Seyfert-1h).  The top axis is the restframe wavelength, the bottom axis the velocity centered in the middle of the OH doublet, and the y-axis is the difference between the flux of the OH doublet and the flux of the continuum, normalized by the factor indicated in the upper-left corner of each panel.  The gray line is the data.  The green shaded region is the integrated line flux used to estimate the EW of the transition.  The velocities are given relative to the blue peak of the doublet, that is v = 0 km/s coincides with the left peak of the doublet when no redshift or blueshift are detected.  The two vertical dashed lines show the galaxy rest wavelength of the doublet.  
    \label{fig:spectra}}
\end{figure*}

The raw spectra were reduced by \citet{ont16} using a custom pipeline which is based on standard procedures from HIPE\footnote{HIPE is a joint development by the $Herschel$ Science Ground Segment Consortium, consisting of ESA, the NASA \textit{Herschel} Science Center, and the HIFI, PACS and SPIRE consortia.} (v13.0.0; \citealt{ott10}), and includes outlier flagging, spectral flat-fielding, regrid of the wavelength sampling, and flux calibration.  Following the standard procedure, most of the sources were observed with the chop-node mode and reduced with the background normalization method, using the off-source spectra to perform the background subtraction and also to correct for the spectral response.  For more on the data reduction, see \citet{ont16}.  A final standard rebinned data cube was produced for each object in the sample.  

Table \ref{tab:sample} contains observational information about the spectra in the sample, including which OH transitions were observed for each galaxy with PACS.  Note that not every galaxy in the sample has observations for all six OH doublets discussed in this paper.  The transitions with the most observations, which provide the largest statistical samples, are OH79, OH119, and OH163.  Figure \ref{fig:spectra} provides a few example spectra from the sample showing the wide variety of OH line profiles observed, and includes absorption, emission, and P-Cygni features.

\subsection{Galaxy Brightness \& IR Line Ratios} \label{subsec:luminosities_and_IR_line_ratios}

To probe the AGN luminosity and strength of the radiation field, we obtain the following luminosities: the intrinsic hard X-ray (2-10 keV) luminosity ($L_{\rm{2-10keV}}$), the [O~\textsc{IV}] 25.89$\mu$m line luminosity ($L_{\rm{[O\ IV]25.89\mu m}}$), and the luminosities of the [Ne~\textsc{V}] 14.32$\mu$m and 24.32$\mu$m emission lines ($L_{\rm{[Ne\ V]14.32\mu m}}$ and $L_{\rm{[Ne\ V]24.32\mu m}}$); and the following IR emission-line ratios: [Ne~\textsc{V}] 24.32$\mu$m/[Ne~\textsc{II}] 12.81$\mu$m, [Ne~\textsc{V}] 14.32$\mu$m/[Ne~\textsc{II}] 12.81$\mu$m, [Ne~\textsc{III}] 15.56$\mu$m/[Ne~\textsc{II}] 12.81$\mu$m, [O~\textsc{IV}] 25.89$\mu$m/[Ne~\textsc{II}] 12.81$\mu$m, and [O~\textsc{IV}] 25.89$\mu$m/[O~\textsc{III}] 88.36$\mu$m.  All of these data were gathered by us in \citet{ont16}.  We use redshift-independent distances (see references within \citealt{ont16}) whenever possible to convert the [O~\textsc{IV}] 25.89$\mu$m, [Ne~\textsc{V}] 24.32$\mu$m, and [Ne~\textsc{V}] 14.32$\mu$m emission-line fluxes and uncertainties to luminosities.  For the rest of the sample, we assume a standard cosmology ($H_0$ = 70 km s$^{-1}$, $\Omega_{\lambda}$ = 0.7, $\Omega_{\rm{M}}$ = 0.3) to calculate luminosity distances when converting from flux to luminosity.

The highly ionized IR lines ([O~\textsc{IV}] 25.89$\mu$m, [Ne~\textsc{V}] 14.32$\mu$m, [Ne~\textsc{V}] 24.32$\mu$m, and [Ne~\textsc{III}] 15.56$\mu$m) serve as a tracer of the harder radiation field produced by the AGN.  The luminosities of these lines, as well as the intrinsic hard X-rays, probe the strength of the AGN.  Comparing the fluxes of these highly ionized lines with the fluxes of lines with lower ionization potential ([Ne~\textsc{II}] 12.81$\mu$m, [O~\textsc{III}] 88.36$\mu$m) traces the relative contribution of an AGN compared to an H II or starburst region.  Therefore, we expect this ratios will be higher for galaxies with a bright AGNs compared with non-active galaxies (see e.g. \citealt{tom08, tom10}).

The X-ray data for $L_{\rm{2-10keV}}$ come from a variety of telescopes.  Chandra and XMM-Newton were the primary telescopes used, but observations from Suzaku, \textit{Swift}/BAT, ASCA and more were also used.  While the observations were not limited to the nuclear region of the galaxies, we assume that the integrated 2-10 keV flux comes only from this central region because it is too energetic for the star formation in the galaxy to produce in significant amounts.

The [O~\textsc{III}] 88.36$\mu$m flux is taken from the central spaxel ($9.4" \times 9.4"$) of the \textit{Herschel}/PACS spectra.  EW(OH) for our sample is also measured using only the central spaxel of the \textit{Herschel}/PACS spectra (see Section \ref{subsec:analysis} for a complete description of how EW(OH) was measured).  All of the other IR emission-lines were observed using the high-resolution module of \textit{Spitzer}/IRS.  [Ne~\textsc{II}] 12.81$\mu$m, [Ne~\textsc{V}] 14.32$\mu$m, and [Ne~\textsc{III}] 15.56$\mu$m were observed using the short-wavelength arm ($4.7" \times 11.3"$).  [Ne~\textsc{V}] 24.32$\mu$m and [O~\textsc{IV}] 25.89$\mu$m were observed using the long-wavelength arm ($11.1" \times 22.3"$).  For the median redshift in our sample, 0.0164, the spatial sizes for the central spaxel of \textit{Herschel}/PACS, the short-wavelength \textit{Spitzer}/IRS arm, and the long-wavelength \textit{Spitzer}/IRS arm are $3.23 \times 3.23$ kpc$^2$, $1.62 \times 3.89$ kpc$^2$, and $3.82 \times 7.67$ kpc$^2$, respectively.  

We assume that the flux for the high-ionization emission-lines ([Ne~\textsc{V}] 14.32$\mu$m, [Ne~\textsc{III}] 15.56$\mu$m, [Ne~\textsc{V}] 24.32$\mu$m, and [O~\textsc{IV}] 25.89$\mu$m) is primarily from the nucleus rather than star formation.  Both the short- and long-wavelength arms of \textit{Spitzer}/IRS capture the entirety of the nucleus, so for these lines it should not matter which arm the fluxes were observed with.  Based on the spatial sizes of the \textit{Herschel}/PACS and \textit{Spitzer}/IRS data, we can capture more of the star formation in the galaxy with [O~\textsc{III}] 88.36$\mu$m compared to [Ne~\textsc{II}] 12.81$\mu$m.  However, both spectra focus on the nucleus and surrounding region of the galaxy and the areas are similar, so the two emission-lines should function as comparable tracers of star formation.

\subsection{IRAS} \label{subsec:IRAS}  

\begin{deluxetable*}{cccc}
\tabletypesize{\scriptsize}
\tablecolumns{4}
\tablecaption{IRAS Data \label{tab:other_data}}
\tablehead{
  \colhead{Object Name} &
  \colhead{IRAS} &
  \colhead{$f_{25\mu\rm{m}}$} &
  \colhead{$f_{60\mu\rm{m}}$} \\
  \colhead{ } & \colhead{Catalog} & \colhead{Flux} & \colhead{Flux}\\
  \colhead{ } & \colhead{ } & \colhead{(Jy)} & \colhead{(Jy)}\\
  \colhead{(1)} & \colhead{(2)} & \colhead{(3)} & \colhead{(4)}}
\startdata
Mrk334 & PSC v2.1 & 1.054 $\pm$ 0.116 & 4.225 $\pm$ 0.511 \\
IRAS00182-7112 & PSC v2.0 & $<$0.250 & 1.300 $\pm$ 0.130 \\
IRAS00198-7926 & PSC v2.1 & 1.210 $\pm$ 0.097 & 3.196 $\pm$ 0.288 \\
NGC253 & PSC v2.1 & 117.100 $\pm$ 4.684 & 758.700 $\pm$ 45.522 \\
Mrk348 & PSC v2.0 & 0.773 $\pm$ 0.062 & 1.440 $\pm$ 0.115 \\
IZw1 & FSC v2.0 & 1.210 $\pm$ 0.121 & 2.240 $\pm$ 0.179 \\
IRAS00521-7054 & PSC v2.1 & 0.837 $\pm$ 0.059 & 1.107 $\pm$ 0.077 \\
ESO541-IG12 & PSC v2.1 & 0.294 $\pm$ 0.047 & 0.718 $\pm$ 0.072 \\
IRAS01003-2238 & PSC v2.0 & 0.555 $\pm$ 0.044 & 2.240 $\pm$ 0.179 \\
NGC454E & FSC v2.0 & 0.417 $\pm$ 0.033 & 1.480 $\pm$ 0.118 \\
\enddata
\tablecomments{
  Col. (1): Target name.
  Col. (2): IRAS Catalog used.
  Col. (3) \& (4): Flux density and uncertainty of the 25$\mu$m and 60$\mu$m lines (Jy).}
\tablecomments{Table \ref{tab:other_data} is published in its entirety in the machine-readable format.
      A portion is shown here for guidance regarding its form and content.}
\end{deluxetable*}

To obtain information about the thermal dust continuum emission for each galaxy in the sample, archival data observed by the \textit{Infrared Astronomical Satellite} (IRAS) were obtained in two bandpasses: 19 $-$ 30$\mu$m and 40 $-$ 80$\mu$m (corresponding to central wavelengths of 25$\mu$m and 60$\mu$m).  Those two bandpasses have an average 10-$\sigma$ sensitivity of 0.65 Jy and 0.85 Jy respectively.  

The ratio of the 25$\mu$m and 60$\mu$m flux densities, $f_{25\mu\rm{m}}$ and $f_{60\mu\rm{m}}$ respectively, are used to estimate the thermal dust continuum emission and act as an IR tracer for AGN strength.  The majority of the 25$\mu$m flux is from warm dust surrounding the AGN, while the 60$\mu$m flux is primarily from cooler dust associated to star formation.  Therefore, a stronger AGN that is more \textquotedblleft dominant" than the host galaxy should have a larger 25$\mu$m/60$\mu$m flux density ratio (hereafter referred to as \textquotedblleft dust temperature"), since the AGN-heated dust is hotter than that of normal spirals.  The ratio for a weak AGN or a non-active galaxy should be much smaller, because the bulk of the IR continuum comes from cooler dust in the host galaxy disk (e.g. \citealt{ede86, vei09, lee11}).  

The preferred IRAS catalog to use is the Point Source Catalog (PSC) v2.1 \citep{hel88}.  For galaxies that are not in that catalog, data are gathered from other IRAS catalogs: PSC v2.0 \citep{ipac86}, the Faint Source Catalog (FSC) v2.0 \citep{mos90}, and the Faint Source Reject Catalog (FSRC; \citealt{mos92}) and other publications \citep{gol88}.  It is important to note that for galaxies with large angular sizes, these catalogs can sometimes exclude the outer parts of the disk.  Because we are focused on the infrared colors of the nuclear region of these galaxies, this limitation of the catalogs is not a concern in this study.  Six galaxies in the sample had no available IRAS data, therefore, data were collected for 172 of the 178 galaxies.  The flux density, uncertainty, and corresponding catalog for each object is indicated in Table \ref{tab:other_data}.  Some of the data were marked as upper limit measurements in the catalogs.  These measurements are identified with a \textquotedblleft$<$" and no uncertainty is reported.

When using dust temperature, there are two main caveats that need to be addressed. First, dwarf galaxies often have high dust temperatures (e.g. \citealt{wal07}). Second, galaxies with extreme dust opacities (e.g. Arp 220 \citealt{ran11}) could show lower IRAS $f_{25\mu\rm{m}}$/$f_{60\mu\rm{m}}$ values, even if an AGN is present. As previously discussed in Section \ref{subsec:PACS}, our sample does not contain any dwarf galaxies and, while slightly favoring IR-luminous galaxies, is very diverse. Therefore, we are not concerned with the bias of dwarf galaxies, and do not expect galaxies with extreme dust opacities to significantly affect our findings.

\subsection{Obscuration Tracers} \label{subsec:silicates_and_balmer_data}

\begin{deluxetable*}{cccccc}
\tabletypesize{\scriptsize}
\tablecolumns{6}
\tablecaption{9.7$\mu$m Silicate \& Balmer Data \label{tab:silicate_and_balmer_data}}
\tablehead{
  \colhead{Object Name} &
  \colhead{Silicate} & 
  \colhead{Ref.} &
  \colhead{H$\alpha$} &
  \colhead{H$\beta$} &
  \colhead{Ref.} \\
  \colhead{ } & \colhead{Strength} & \colhead{ } & \colhead{(10$^{-14}$erg/s/cm$^2$)} & \colhead{(10$^{-14}$erg/s/cm$^2$)} & \colhead{ } \\
  \colhead{(1)} & \colhead{(2)} & \colhead{(3)} & \colhead{(4)} & \colhead{(5)} & \colhead{(6)}}
\startdata
Mrk334 &  &  & 28.35 & 4.48 & MAL17,OST93 \\
IRAS00182-7112 &  &  &  &  &  \\
IRAS00198-7926 &  &  &  &  &  \\
NGC253 &  &  &  &  &  \\
Mrk348 & $-$0.333 $\pm$ 0.017 & WU09 & 11.96 & 2.80 & MAL86,KOS78 \\
IZw1 & 0.284 $\pm$ 0.014 & WU09 & 74.80 & 31.00 & MAL17,OKE79,VER04 \\
IRAS00521-7054 &  &  &  &  &  \\
ESO541-IG12 &  &  & 1.38 & 0.32 & MAL17 \\
IRAS01003-2238 & $-$0.77 $\pm$ 0.08 & FAR13 &  &  &  \\
NGC454E & $-$0.38 $\pm$ 0.16 & HER15 &  &  &  \\
\enddata
\tablecomments{
  Col. (1): Target name.
  Col. (2): Strength of the silicate absorption feature where a negative value indicates an absorption feature, and a positive value indicates emission (as defined in \citet{spo07}).
  Col. (3): Reference for the 9.7$\mu$m silicate data.  ALO16: \citet{alo16}, FAR13: \citet{far13}, GM15: \citet{gm15}, HER15: \citet{her15}, WU09: \citet{wu09}.  
  Col. (4): Line flux of the H$\alpha$ line (in units of 10$^{-14}$erg/s/cm$^2$).
  Col. (5): Line flux of the H$\beta$ line (in units of 10$^{-14}$erg/s/cm$^2$).
  Col. (6): References for the H$\alpha$ and H$\beta$ data. COH81: \citet{coh81}, CON12: \citet{con12}, DIA88: \citet{dia88}, DUR88: \citet{dur88}, ERK97: \citet{erk97}, GIL10: \citet{gil10}, GOO83: \citet{goo83}, HO93: \citet{ho93}, HO97: \citet{ho97}, KOS78: \citet{kos78}, KRA94: \citet{kra94}, MAL86: \citet{mal86}, MAL17: \citet{mal17}, MOR88: \citet{mor88}, MOU10: \citet{mou10}, OKE79: \citet{oke79}, OST75: \citet{ost75}, OST81: \citet{ost81}, OST83: \citet{ost83}, OST93: \citet{ost93}, PHI83: \citet{phi83}, SHU80: \citet{shu80}, SHU81: \citet{shu81}, VAC97: \citet{vac97}, VEI95: \citet{vei95}, VER04: \citet{ver04}, WIN92: \citet{win92}.} 
\tablecomments{Table \ref{tab:silicate_and_balmer_data} is published in its entirety in the machine-readable format.
      A portion is shown here for guidance regarding its form and content.}
\end{deluxetable*}

Because we expect that the warm molecular gas traced by the OH doublets should be associated with dusty embedded regions, we include in our study the observations for the following dust absorption tracers: the 9.7$\mu$m silicate line and the Balmer decrement.  The observations for the silicate, H$\alpha$, and H$\beta$ lines come from a variety of sources in the literature.  The 9.7$\mu$m silicate line was always measured using the low-resolution module of \textit{Spitzer}/IRS with an aperture of $\sim$3.6", while the optical Balmer lines come from a variety of sources, most of which are ground-based slit spectroscopy (typically $\sim$1" and limited by seeing).  

The strength of the silicate feature is quantified using the method derived by \citet{spo07}:

\begin{equation}
S_{\textup{sil}} = \textup{ln}\frac{f_{\textup{obs}}(9.7\mu\textup{m})}{f_{\textup{cont}}(9.7\mu\textup{m})}  \\
\end{equation}
Where $f_{\rm{obs}}$ is the observed flux density of the 9.7$\mu$m silicate line and $f_{\rm{cont}}$ is the continuum flux density at 9.7$\mu$m.  $S_{\rm{sil}}$ can be interpreted as an estimate of the optical depth for silicate absorption features.

Table \ref{tab:silicate_and_balmer_data} gives the strength of the silicate absorption feature, the Balmer fluxes, and the sources that these measurements come from.  There is some overlap between our sample and those from \citet{spo13}, \citet{vei13}, and \citet{sto16} (OH119) and \citet{gon15} (OH65), which find correlations between the silicate feature and these doublets.  We will test if this relationship exists with our sample for OH65 and OH119, and if the other OH doublets follow a similar trend.

\begin{deluxetable*}{ccccccc}
\tabletypesize{\scriptsize}
\tablecolumns{7}
\tablecaption{Derived OH Integrated Flux Values \label{tab:integrated_oh_fluxes}}
\tablehead{
  \colhead{Object Name} &
  \colhead{OH65} & \colhead{OH71} & \colhead{OH79} & \colhead{OH84} & \colhead{OH119} & \colhead{OH163} \\
    \colhead{ } & \colhead{\tiny{(10$^{-14}$erg/s/cm$^2$)}} & \colhead{\tiny{(10$^{-14}$erg/s/cm$^2$)}} & \colhead{\tiny{(10$^{-14}$erg/s/cm$^2$)}} & \colhead{\tiny{(10$^{-14}$erg/s/cm$^2$)}} & \colhead{\tiny{(10$^{-14}$erg/s/cm$^2$)}} & \colhead{\tiny{(10$^{-14}$erg/s/cm$^2$)}} \\
  \colhead{(1)} & \colhead{(2)} & \colhead{(3)} & \colhead{(4)} & \colhead{(5)} & \colhead{(6)} & \colhead{(7)}}
\startdata
Mrk334 &  &  & $<-$4.78 &  &  &  \\
IRAS00182-7112 &  &  & $<-$1.01 & $<-$1.43 & $-$1.09 $\pm$ 0.31 & \\
IRAS00198-7926 &  &  & 3.41 $\pm$ 1.90 &  &  & \\
NGC253 & $-$49.08 $\pm$ 28.76 & $<-$250.82 & $-$102.72 $\pm$ 79.66 & $-$168.79 $\pm$ 69.64 & $-$1193.81 $\pm$ 21.93 & 125.40 $\pm$ 18.35 \\
Mrk348 &  &  &  & $<-$0.94 &  & \\
IZw1 & $<-$21.64 &  & 3.57 $\pm$ 0.72 &  & 2.02 $\pm$ 0.46 & 1.31 $\pm$ 0.57 \\
IRAS00521-7054 &  &  & $<-$5.29 &  &  & \\
ESO541-IG12 &  &  & $<-$2.32 &  &  & \\
IRAS01003-2238 &  &  & $<-$3.38 &  & $-$3.15 $\pm$ 1.06 & \\
NGC454E &  &  & $<-$3.13 &  &  & \\
\enddata
\tablecomments{
  Col. (1): Target name.
  Col. (2): Integrated flux value and uncertainty of the OH65 doublet.
  Col. (3): Integrated flux value and uncertainty of the OH71 doublet.
  Col. (4): Integrated flux value and uncertainty of the OH79 doublet.
  Col. (5): Integrated flux value and uncertainty of the OH84 doublet.
  Col. (6): Integrated flux value and uncertainty of the OH119 doublet.
  Col. (7): Integrated flux value and uncertainty of the OH163 doublet.  Note that if an integrated flux measurement does not have an uncertainty value, it is an upper limit observation.  If there is an uncertainty, it is a detection.}
\tablecomments{Table \ref{tab:integrated_oh_fluxes} is published in its entirety in the machine-readable format.
      A portion is shown here for guidance regarding its form and content.}
\end{deluxetable*}

\begin{deluxetable*}{cccccccc}
\tabletypesize{\scriptsize}
\tablecolumns{8}
\tablecaption{Derived EW(OH) Values \label{tab:data}}
\tablehead{
  \colhead{Object Name} &
  \colhead{Classification} &
  \colhead{OH65} & \colhead{OH71} & \colhead{OH79} & \colhead{OH84} & \colhead{OH119} & \colhead{OH163} \\
    \colhead{ } & \colhead{ } & \colhead{($\mu$m)} & \colhead{($\mu$m)} & \colhead{($\mu$m)} & \colhead{($\mu$m)} & \colhead{($\mu$m)} & \colhead{($\mu$m)} \\
  \colhead{(1)} & \colhead{(2)} & \colhead{(3)} & \colhead{(4)} & \colhead{(5)} & \colhead{(6)} & \colhead{(7)} & \colhead{(8)}}
\startdata
Mrk334 & Sy-1.8 &  &  & $<$0.022 &  &  &  \\
IRAS00182-7112 & Sy-2 &  &  & $<$0.016 & $<$0.030 & $-$0.078 $\pm$ 0.022 & \\
IRAS00198-7926 & Sy-2 &  &  & 0.028 $\pm$ 0.015 &  &  & \\
NGC253 & Starburst & $-$0.001 $\pm$ 0.001 & $<$0.006 & $-$0.003 $\pm$ 0.002 & $-$0.006 $\pm$ 0.002 & $-$0.078 $\pm$ 0.002 & 0.028 $\pm$ 0.004 \\
Mrk348 & Sy-1h &  &  &  & $<$0.045 &  & \\
IZw1 & Sy-1n & $<$0.190 &  & 0.033 $\pm$ 0.007 &  & 0.051 $\pm$ 0.012 & 0.087 $\pm$ 0.038 \\
IRAS00521-7054 & Sy-1h &  &  & $<$0.131 &  &  & \\
ESO541-IG12 & Sy-2 &  &  & $<$0.079 &  &  & \\
IRAS01003-2238 & Sy-2 &  &  & $<$0.042 &  & $-$0.154 $\pm$ 0.052 & \\
NGC454E & Sy-2 &  &  & $<$0.109 &  &  & \\
\enddata
\tablecomments{
  Col. (1): Target name.
  Col. (2): Optical spectral type.
  Col. (3): EW and uncertainty of the OH65 doublet.
  Col. (4): EW and uncertainty of the OH71 doublet.
  Col. (5): EW and uncertainty of the OH79 doublet.
  Col. (6): EW and uncertainty of the OH84 doublet.
  Col. (7): EW and uncertainty of the OH119 doublet.
  Col. (8): EW and uncertainty of the OH163 doublet.  Note that if an EW measurement does not have an uncertainty value, it is an upper limit observation.  If there is an uncertainty, it is a detection.}
\tablecomments{Table \ref{tab:data} is published in its entirety in the machine-readable format.
      A portion is shown here for guidance regarding its form and content.}
\end{deluxetable*}

\section{Analysis \& Results}\label{sec:analysis_and_results}

In this section, we first describe how we obtained our EW measurements from the archival PACS datacubes.  Then, to uncover which observed galaxy properties may predict emission/absorption for the OH doublets in our sample, we investigate possible correlations between these EW values and galaxy observables such as AGN luminosity, radiation field hardness, dust temperature traced by the IRAS $f_{25\mu\rm{m}}$/$f_{60\mu\rm{m}}$ ratio, and dust extinction.  A positive EW value indicates that the transition is in emission while a negative value indicates absorption.

\subsection{Measuring EW}\label{subsec:analysis}

From the reduced data cubes, spectra from the central spaxel ($9.4" \times 9.4"$ or $3.23 \times 3.23$ kpc$^2$ at the median redshift of our sample, 0.0164) were extracted.  Focusing on the central spaxel will magnify the effects of a possible AGN component on the OH transitions.  We used Python scripts developed by us in \citet{ont16} to measure the flux, estimate the continuum level for each OH transition, and obtain the EWs.  The continuum was estimated using a 1-D polynomial fit to the spectral channels located in the red and blue wings next to the OH line.  The integrated flux values with uncertainties are shown in Table \ref{tab:integrated_oh_fluxes}.

We do see complex kinematics of the OH gas for some of the galaxies in the sample. Typical cases include the doublets that display P-Cygni profiles, which have both positive and negative components (see Figure \ref{fig:spectra}).  Other line profiles have prominent broad wings, which could be signatures of outflows.  Also, some measurements are blended with weak neighboring spectral lines.  The only case that the numerical integration method gives an unreliable EW measurement is with P-Cygni features, therefore, we remove these observations when looking for for correlations between EW(OH) and galaxy observables.  We only observe P-Cygni profiles in a minority of galaxies (see Section \ref{subsec:p_cyg} below for more details), so this does not notably reduce our sample size.  

The EW and the uncertainty of these OH transitions are derived, and Table \ref{tab:data} shows the EW measurements with uncertainties for each object.  For some galaxies, only the continuum was detected, with no OH emission or absorption above or below the noise level, respectively.  In this case, the continuum flux density ($C$) was used to estimate an upper limit for the EW as $3 \sigma /C$.  $\sigma$ corresponds to the RMS of the continuum emission along the spectral range observed.  The upper limit EW(OH) measurements are identified with a \textquotedblleft$<$" in Table \ref{tab:data}.

\subsubsection{Caveat on EW(OH) Emission \& Absorption}

In this paper, we focus on the study of the most fundamental observed properties of OH spectra in galaxies.  The first most basic measurement is whether an OH doublet is mostly in emission or absorption (i.e. is the sign of EW(OH) positive or negative).  This simple observable is expected to indicate the importance of a) excitation of higher energy levels in the OH molecules versus b) higher column densities of molecules in their lower energy states.  Galaxies in which a) is more important should show more OH emission lines, whereas those in which b) is more important should show stronger absorption.  

Once the sign of the EW(OH) is measured, the second most important quantity is its strength.  We therefore search for statistical trends in the sign and strength of EW(OH) with other galaxy properties, which correlate with molecular excitation or optical depth.  We now consider both of these correlations in turn.

Even though absorption and emission of a spectral line have different physical origins, we intentionally include the two different types of OH line profiles in the same analysis when looking for trends in the sample.  This is because our goal is to find global trends with observables that could affect EW(OH) (e.g. total luminosity, hardness of the radiation field, and extinction) that all galaxies in the local universe follow, including if EW(OH) changes from tending to be in emission to absorption (or vice versa).

\subsection{Comments on the Sample}

In this section, we identify some trends observed for EW(OH) in the sample data.  As noted in Section \ref{subsec:analysis}, the EW(OH) values can be found in Table \ref{tab:data}.

\subsubsection{Doublets Favoring Absorption} 

While it has only ten detections, OH65 is found almost exclusively in absorption, with only one detection (10\% of the sample) showing emission.  The galaxy with EW(OH65) emission is NGC 2146, which is a non-active galaxy, but is a bright LIRG.  For OH71, all five detections are seen in absorption.  This could be due to the fact that these transitions originate at the highest energies (the lower energy levels of both transitions are between 300-400 K) on the OH Grotrian diagram.  

Although not to the same degree as the OH65 and OH71 doublets, OH84 is also primarily seen in absorption (85\% of detections).  The two galaxies seen in emission are ESO 103-G35 and NGC 1068, which are both optically classified as Sefyert-2 galaxies.

\subsubsection{OH163 in Emission} \label{subsec:oh163}

One important finding is that OH163 is always detected in emission (25 galaxies).  No other OH doublet in this study is found exclusively in emission.  OH79 strongly favors emission (89\% of detections), but unlike OH163, it can also be found in absorption.

\subsubsection{OH119} \label{subsec:oh119}

OH119 is the most diverse doublet in the sample, showing the largest variety in profiles, and unlike the other doublets discussed previously, the number of galaxies seen in emission and absorption is approximately equal.  When we investigate possible correlations between EW(OH) and various galaxy observables in the rest of this study, the OH119 doublet is the most discussed transition from the sample because it has the highest number of detections, which makes the results the most statistically significant.

\subsubsection{Relative EW Strengths between Transitions}

Not including upper limits, for the doublets that are found primarily in absorption $-$ OH65, OH71, and OH84 $-$ OH65 and OH71 have the weakest detected EW values (strongest EW absorption = $-$0.028$\mu$m and $-$0.017$\mu$m respectively) with OH84 (strongest EW absorption = $-$0.056$\mu$m) being two or three times stronger.  For galaxies with measurements of both OH65 and OH84, the EW of the latter is also about twice as large.  However, this is based only on 7 galaxies that have detections for both doublets.  These relative strengths of the observed OH doublets could be explained by the OH Grotrian diagram shown in Figure \ref{fig:energy_diagram}.  OH65 and OH71 are the highest energy transitions in the sample, making it harder for the lower levels to be populated.  OH84 is the level immediately below OH65, so OH84 could be stronger because the lower energy level corresponds to a lower excitation temperature and therefore is easier to populate than OH65 and OH71.

OH119, which is observed in both emission and absorption with regularity, has maximum absorption and emission values of $-$0.163$\mu$m and 0.103$\mu$m respectively.  This maximum absorption for OH119 is about an order of magnitude stronger than OH65 and OH71, and about twice as strong as OH84.  For OH79, which favors emission, the maximum absorption is very weak ($-$0.027$\mu$m) and is about even strength with OH65 and OH71.  The emission for OH79 is much stronger (max EW emission = 0.242$\mu$m), which is about twice as strong as OH119.  OH163, which is only seen in emission, has comparable emission strength to that of OH119.

\subsubsection{P-Cygni Features} \label{subsec:p_cyg}

Some of our galaxies contain one or more OH transitions with a P-Cygni profile, a clear indication that the galaxy has a molecular outflow moving out from the AGN component.  The P-Cygni features are identified by eye, and it was found that 19 galaxies in the sample show P-Cygni features in one or more OH transitions.  Out of these 19 galaxies, five (IRAS 01003-2238, IRAS 17208-0014, IRAS 23365+3604, Mrk 273, and NGC 6240) have two OH transitions that show a P-Cygni feature and one (Mrk 231) contains three transitions that display it.  Many of these P-Cygni features have been published in previous studies, with about half of them having been identified in \citet{vei13}.  Table \ref{tab:pcyg} shows which OH transitions for which galaxies display a P-Cygni feature.  Table \ref{tab:pcyg_stats} shows the percentage of observations that show a P-Cygni feature for each OH transition.  

\startlongtable
\begin{deluxetable}{cc}
\tabletypesize{\scriptsize}
\tablecolumns{2}
\tablecaption{List of Objects Showing P-Cygni Features \label{tab:pcyg}}
\tablehead{
  \colhead{Object Name} &
  \colhead{OH Transitions with P-Cygni Feature}\\
  \colhead{ } & \colhead{($\mu$m)}\\
  \colhead{(1)} & \colhead{(2)}}
\startdata
 Circinus & 119* \\
 IRAS 01003-2238 & 79, 119 \\
 IRAS 05189-2524 & 119 \\
 IRAS 12071-0444 & 119 \\
 IRAS 13120-5453 & 79 \\
 IRAS 14394+5332 & 119 \\
 IRAS 15462-0450 & 119 \\
 IRAS 17208-0014 & 79 \\
 IRAS 19254-7245 & 119 \\
 IRAS 23365+3604 & 79, 119 \\
 Mrk 231 & 71, 79, 119 \\
 Mrk 273 & 79, 119 \\
 Mrk 848B & 79 \\
 NGC 253 & 79 \\
 NGC 1266 & 79 \\
 NGC 3079 & 119 \\
 NGC 5728 & 119 \\
 NGC 6240 & 84, 119 \\
 NGC 7674 & 79 \\
\enddata
\tablecomments{
  Col. (1): Name of the object.
  Col. (2): OH transitions that show a P-Cygni feature for the object.
  Note that a * indicates that it is a reverse P-Cygni feature.}
\end{deluxetable}

The two tables show that OH79 and OH119 have the highest number of P-Cygni observations as well as the highest percentage.  We find one P-Cygni feature in the OH71 and OH84 doublets (Mrk 231 and NGC 6240 respectively).  OH65 and OH163 do not show any P-Cygni features.  

Circinus is a unique galaxy in our sample, because it is the only one to show a reverse P-Cygni profile in the nuclear spectrum.  A reverse P-Cygni profile indicates that there is an inflow instead of an outflow of molecular gas.

\startlongtable
\begin{deluxetable}{ccc}
\tabletypesize{\scriptsize}
\tablecolumns{3}
\tablecaption{Percent of P-Cygni Features Found for each OH Transition \label{tab:pcyg_stats}}
\tablehead{
  \colhead{OH Transition} & 
  \colhead{Number of P-Cygni} & 
  \colhead{Percent that show a}\\
  \colhead{ } & 
  \colhead{Features Observed} & 
  \colhead{P-Cygni Feature} \\
  \colhead{($\mu$m)} & \colhead{ } & \colhead{ }\\
  \colhead{(1)} & \colhead{(2)} & \colhead{(3)}}
\startdata
 65 & 0 out of 23 & 0\% \\
 71 & 1 out of 13 & 7.7\% \\
 79 & 10 out of 152 & 6.6\% \\
 84 & 1 out of 20 & 5.0\% \\
 119 & 13 out of 86 & 15.1\% \\
 163 & 0 out of 29 & 0\% \\
\enddata
\tablecomments{
  Col. (1): OH transition.
  Col. (2): Number of galaxies with P-Cygni features observed.
  Col. (3): Percent of observations that displayed a P-Cygni or reverse P-Cygni feature.}
\end{deluxetable}

We note that \citet{gon17a} identifies the PACS observation for OH84 in IRAS 23365+3604 as a P-Cygni feature, but we do not.  The key reason for this difference is that \citet{gon17a} uses a 3 $\times$ 3 spaxel area for their spectra, while we only use the central spaxel.  This means that if the P-Cygni feature is tracing a molecular outflow in the galaxy, it has left the central region of the galaxy.

As discussed above in Section \ref{subsec:analysis}, the EW measurements for P-Cygni features are unreliable.  Therefore, we will not include the P-Cygni lines reported in Table \ref{tab:pcyg} in the rest of this section when we investigate correlations between EW(OH) and galaxy observables.

\subsection{EW(OH) vs. AGN Luminosity and Radiation Field Hardness}\label{subsec:width_vs_type}

The far-IR continuum is typically dominated by thermal emission from dust grains, which can be heated either by absorbing UV/optical emission from young stars or an AGN.  In this paper, we use several observables to quantify the relative importance of star formation and non-stellar nuclear activity.  Both the column density and temperature of the dust may differ depending on the strength of these mechanisms in the central region of the host galaxy.  We now investigate which of the observables introduced in Sections \ref{subsec:luminosities_and_IR_line_ratios}, \ref{subsec:IRAS}, and \ref{subsec:silicates_and_balmer_data} may correlate with EW(OH).  A positive correlation mainly tells us that increasing a physical parameter makes the corresponding OH doublet more likely to appear in emission.  

For the rest of Section \ref{sec:analysis_and_results}, we use the following guidelines when investigating possible correlations between EW(OH) and the galaxy observables.  We only consider detections for both EW(OH) and the host galaxy properties we look to correlate EW(OH) with.  Upper limit observations are excluded when looking for trends between EW(OH) and these observables so that the limits do not hide any correlations.  Also, a minimum threshold of 8 galaxies is adopted when looking for correlations between EW(OH) and the galaxy observables.  We use the Spearman rank-order correlation coefficient, $r$, and p-value to judge the strength of possible relationships.  A ``significant correlation'' is identified as having $|r|$ $\geq$ 0.5 and p $<$ 0.01 and a ``weak correlation'' is identified as having $|r|$ $\geq$ 0.5 and 0.01 $<$ p $<$ 0.05.  When p $<$ 0.05 but $|r|$ $<$ 0.5, this indicates that there is a lot of scatter in the relationship, and we therefore deem these observables to be uncertain predictors of EW(OH).

\subsubsection{Line Ratios} \label{subsec:line_ratios}

Optical emission may be heavily absorbed by dust, therefore, we turn to IR tracers.  Mid- and far-IR spectroscopy provides measures of the emission produced by accretion onto a supermassive black hole in AGNs and the emission produced by star-formation (e.g. \citealt{spi92}).  Using ratios of a high-ionization potential emission-line (AGN-dominated) to a low-ionized emission-line (H II region or starburst-dominated) can probe the relative contribution of both sources in a galaxy (e.g. \citealt{tom08, tom10, spi15, ont16}).  The larger the ratio means that a galaxy has a larger AGN contribution, while a smaller ratio would indicate that there is little or no AGN activity (starburst-dominated galaxy).  

We investigate the following emission-line ratios against EW(OH): [Ne~\textsc{V}] 24.32$\mu$m/[Ne~\textsc{II}] 12.81$\mu$m, [Ne~\textsc{V}] 14.32$\mu$m/[Ne~\textsc{II}] 12.81$\mu$m, [Ne~\textsc{III}] 15.56$\mu$m/[Ne~\textsc{II}] 12.81$\mu$m, [O~\textsc{IV}] 25.89$\mu$m/[Ne~\textsc{II}] 12.81$\mu$m, and [O~\textsc{IV}] 25.89$\mu$m/[O~\textsc{III}] 88.36$\mu$m.  These ratios are set up so that the emission-line in the numerator is more highly ionized, indicating that it arises in a warmer (AGN) environment.  The emission-line in the denominator has lower ionization potential.  Therefore, we expect AGN dominated galaxies to have higher IR emission-line ratios compared to obscured AGNs and non-active galaxies.  

\begin{figure*}[t!]
    \centering
    \includegraphics[scale=0.36]{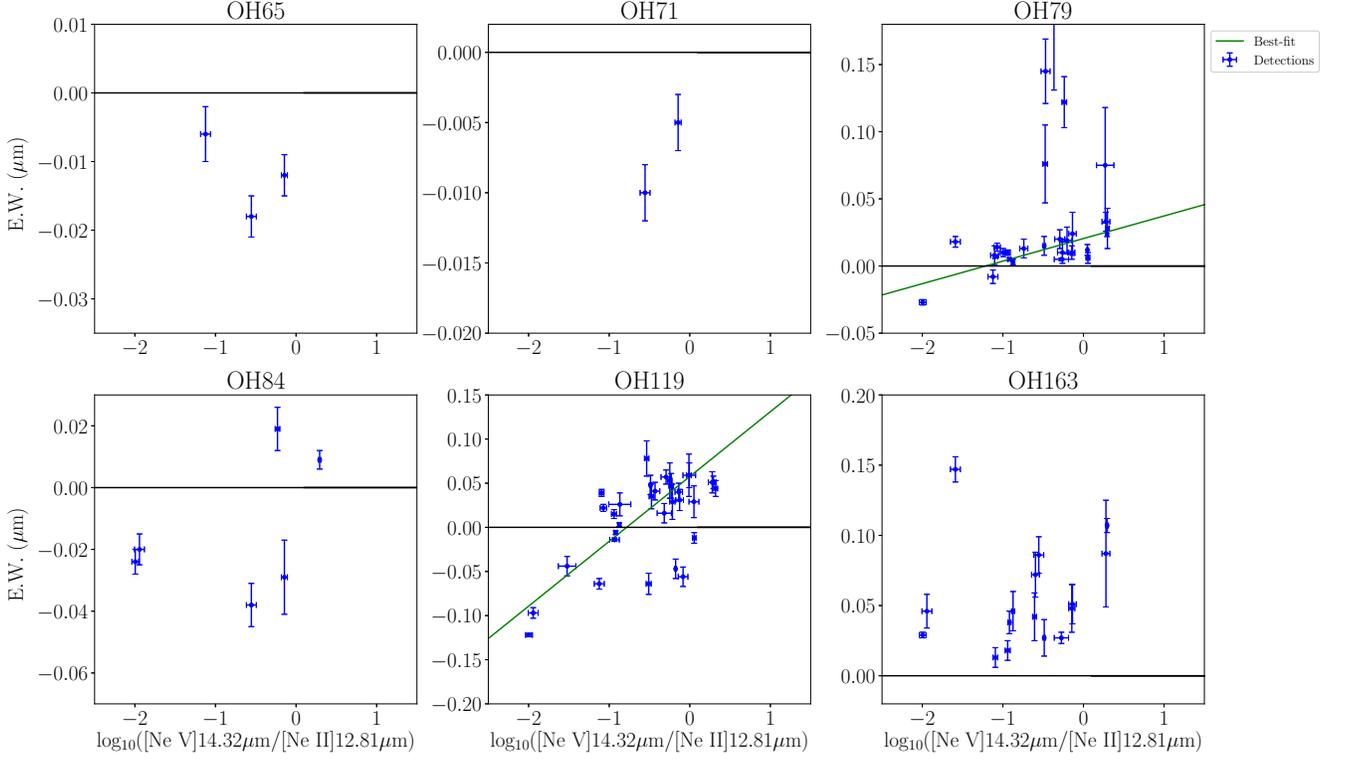}
    \caption{EW of the OH doublet vs. the logarithm of the [Ne~\textsc{V}] 14.32$\mu$m/[Ne~\textsc{II}] 12.81$\mu$m emission-line ratio.  Galaxies must have detections in EW(OH), [Ne~\textsc{V}] 14.32$\mu$m, and [Ne~\textsc{II}] 12.81$\mu$m to be included in the plot and fits.  The weighted log-linear fits, using the parameters given in Table \ref{tab:line_ratio_flux_stats}, are shown for fits with p $<$ 0.05.  We do not show the fit for slopes with a p-value $>$ 0.05.  For OH79, MCG-01-24-012 is an outlier (EW = 0.242$\mu$m), and is cut off to better show the rest of the data.}
    \label{fig:nev14_neii_plot}
\end{figure*}

\begin{figure*}[t!]
    \centering
    \includegraphics[scale=0.36]{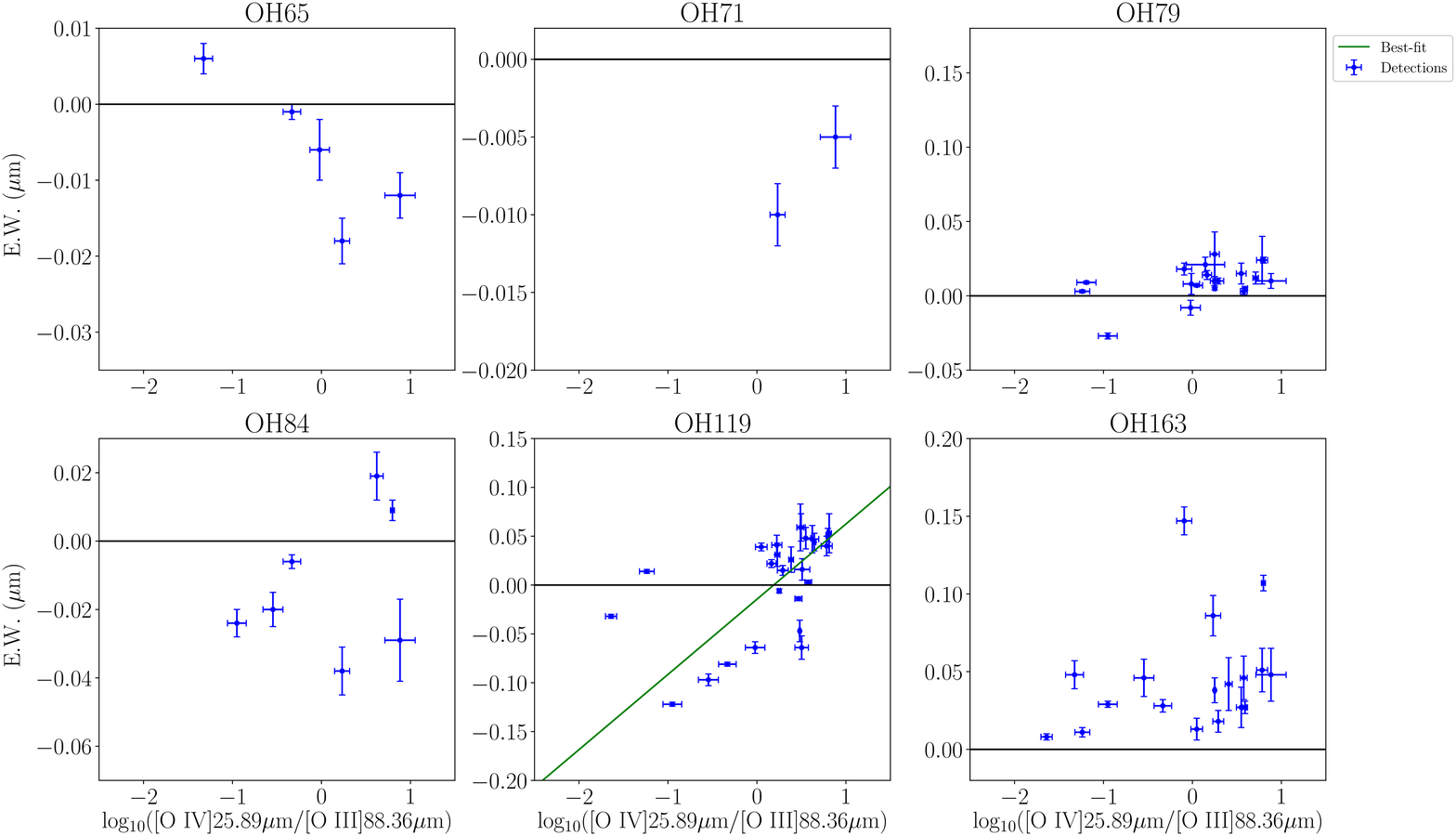}
    \caption{EW of the OH doublet vs. the logarithm of the [O~\textsc{IV}] 25.89$\mu$m/[O~\textsc{III}] 88.36$\mu$m emission-line ratio.  Galaxies must have detections in EW(OH), [O~\textsc{IV}] 25.89$\mu$m, and [O~\textsc{III}] 88.36$\mu$m to be included in the plot and fits.  The weighted log-linear fits, using the parameters given in Table \ref{tab:line_ratio_flux_stats}, are shown for fits with p $<$ 0.05.  We do not show the fit for slopes with a p-value $>$ 0.05.}
    \label{fig:oiv_oiii_plot}
\end{figure*}

An orthogonal distance regression using a modified trust-region Levenberg-Marquardt-type algorithm was used to obtain a first degree polynomial fit for the data of each OH doublet.  The fit is weighted using the uncertainties from both EW(OH) and the emission-line ratios.  To obtain the uncertainty for the line ratios, we add the uncertainties of the emission-lines in quadrature.  As mentioned above, we exclude any galaxies for which either EW(OH) or the emission-line ratio is an upper limit observation.  We estimated a fit for each EW(OH) against each emission-line ratio individually.  Table \ref{tab:line_ratio_flux_stats} in the Appendix gives the results for the log-linear fits.  Plots for EW(OH) vs. [Ne~\textsc{V}] 14.32$\mu$m/[Ne~\textsc{II}] 12.81$\mu$m and [O~\textsc{IV}] 25.89$\mu$m/[O~\textsc{III}] 88.36$\mu$m are shown in Figures \ref{fig:nev14_neii_plot} and \ref{fig:oiv_oiii_plot} respectively.  There is one panel for each OH doublet, and the fits from Table \ref{tab:line_ratio_flux_stats} that have p $<$ 0.05 are show in the Figures as a solid green line.  Figures for EW(OH) vs. the other three line ratios ([Ne~\textsc{V}] 24.32$\mu$m/[Ne~\textsc{II}] 12.81$\mu$m, [O~\textsc{IV}] 25.89$\mu$m/[Ne~\textsc{II}] 12.81$\mu$m, and [Ne~\textsc{III}] 15.56$\mu$m/[Ne~\textsc{II}] 12.81$\mu$m) are available as supplementary material in the online version of the paper.

We find that the OH119 doublet is the most significantly affected by changes to the emission-line ratios.  Table \ref{tab:line_ratio_flux_stats} shows that EW(OH119) has a significant correlation ($r$ $\geq$ 0.5 and p $<$ 0.01) with log([Ne~\textsc{V}] 14.32$\mu$m/[Ne~\textsc{II}] 12.81$\mu$m), log([Ne~\textsc{III}] 15.56$\mu$m/[Ne~\textsc{II}] 12.81$\mu$m), and log([O~\textsc{IV}] 25.89$\mu$m/[O~\textsc{III}] 88.36$\mu$m).  The latter two line ratios are the strongest predictors of EW(OH119), with $r$ $>$ 0.60.  Due to the large amount of scatter ($r$ = 0.45, p = 0.005), log([O~\textsc{IV}] 25.89$\mu$m/[Ne~\textsc{II}] 12.81$\mu$m) is shown to be an uncertain predictor of EW(OH119).  OH119 is more likely to be in absorption for low emission-line ratios, and transitions to emission as the high-ionization IR lines become stronger.  Since high-ionization lines are dominated by the AGN emission, we can associate an enhanced OH119 emission with the presence of an AGN.

While already significant ($r$ = 0.52, p = 0.0025), it is worth noting that the strength of the correlation between EW(OH119) and log([Ne~\textsc{V}] 14.32$\mu$m/[Ne~\textsc{II}] 12.81$\mu$m) greatly increases in strength ($r$ = 0.71, p = 0.00003) when we remove IRAS 12514+1027, NGC 5506, PKS 2048-57, and NGC 7172 from the fit.  These four galaxies are identified as outliers because EW(OH119) is in absorption, while the rest of the sample is in emission at log([Ne~\textsc{V}] 14.32$\mu$m/[Ne~\textsc{II}] 12.81$\mu$m) $\gtrsim -$0.5.  Removing the same four galaxies also makes log([O~\textsc{IV}] 25.89$\mu$m/[Ne~\textsc{II}] 12.81$\mu$m) a much better predictor of EW(OH119) as well ($r$ = 0.58, p = 0.0004).  Based on our methodology outlined in Section \ref{subsec:width_vs_type}, without these outliers, EW(OH119) vs. log([O~\textsc{IV}] 25.89$\mu$m/[Ne~\textsc{II}] 12.81$\mu$m) displays a significant correlation.  One explanation for these four outlier galaxies could be that the molecular gas seen in OH does not react instantaneously to the AGN activity, so these few cases out of the correlation might be recent AGN sources that need some Myrs to drive OH119 into emission (i.e. into the correlation).

While to a lesser degree, EW(OH79) and EW(OH163) are also correlated with the IR emission-lines ratios.  EW(OH79) shows a significant correlation ($r$ = 0.50, p = 0.007) with log([Ne~\textsc{V}] 14.32$\mu$m/[Ne~\textsc{II}] 12.81$\mu$m).  Due to low p-values but $r$ values of 0.46 and 0.37, respectively, log([Ne~\textsc{III}] 15.56$\mu$m/[Ne~\textsc{II}] 12.81$\mu$m) and log([O~\textsc{IV}] 25.89$\mu$m/[Ne~\textsc{II}] 12.81$\mu$m) are found to be uncertain predictors of EW(OH79).  The $r$ value for the EW(OH79) relationships is always smaller compared to that of EW(OH119), indicating that while the IR line ratios can predict EW(OH79), it is not as strong.  Despite the increase in scatter, this doublet displays strong emission when the ratio of the high ionization to low ionization lines is largest, which indicates that a more highly ionized galaxy (due to the presence of an AGN) will contain OH79 emission, similar to OH119.  At lower flux ratio values, OH79 primarily shows weak emission, with only a few galaxies in absorption.  This is different from OH119, which displayed deep absorption when the low ionization line was significantly stronger.  There are four galaxies, IC 1816, MCG-01-24-012, Mrk 705, Mrk 883, with extreme OH79 emission ($\rm{EW}>+0.10$) compared to the rest of the sample; however, removing these objects does not significantly improve the strength of any of the three fits.

EW(OH163) displays a weak correlation ($r$ = 0.69, p = 0.02) with log([Ne~\textsc{V}] 24.32$\mu$m/[Ne~\textsc{II}] 12.81$\mu$m).  Also, log([Ne~\textsc{III}] 15.56$\mu$m/[Ne~\textsc{II}] 12.81$\mu$m) is found to be an uncertain predictor of the doublet ($r$ = 0.43, p = 0.03).  In both of these relationships, stronger OH163 emission is found in the more ionized galaxies, similar to both OH119 and OH79.  OH163 is more likely to display weaker emission at lower flux ratio values.  There are no obvious outliers in the relationship with log([Ne~\textsc{III}] 15.56$\mu$m/[Ne~\textsc{II}] 12.81$\mu$m); therefore, log([Ne~\textsc{V}] 24.32$\mu$m/[Ne~\textsc{II}] 12.81$\mu$m) is the better predictor of EW(OH163).

Based on the criteria outlined in Section \ref{subsec:width_vs_type}, no correlations are found between EW(OH65), EW(OH71), and EW(OH84) with any of the emission-line ratios.  This is primarily due to a lack of observations.  There could be possible correlations between EW(OH65) and both log([O~\textsc{IV}] 25.89$\mu$m/[Ne~\textsc{II}] 12.81$\mu$m) and log([O~\textsc{IV}] 25.89$\mu$m/[O~\textsc{III}] 88.36$\mu$m) as $r$ = $-$0.90 and p = 0.04 for both relationships.  However, there are only five galaxies in both of these samples, which are too few to conclude if these correlations exist.  If the EW(OH65) correlations are real, this would imply that a warm gas with a significant fraction of molecules at $\sim$300 K (see Figure \ref{fig:energy_diagram}) is needed to drive OH65 absorption.  A relatively luminous source and/or a strong radiation field is needed to achieve temperatures this high.

It is important to note that the normal spiral galaxy M83, which has no indications of AGN activity any of these observables, shows OH79 and OH119 weakly in emission.  As discussed, none of the correlations are perfect, and the occasional outlier is expected.  Therefore, we caution that some of the generalizations that we draw from the full sample may not necessarily apply to every galaxy.  Our physical interpretation of the OH transitions may not fully explain the observed OH doublets in every galaxy.

\begin{figure*}[t!]
    \centering
    \includegraphics[scale=0.36]{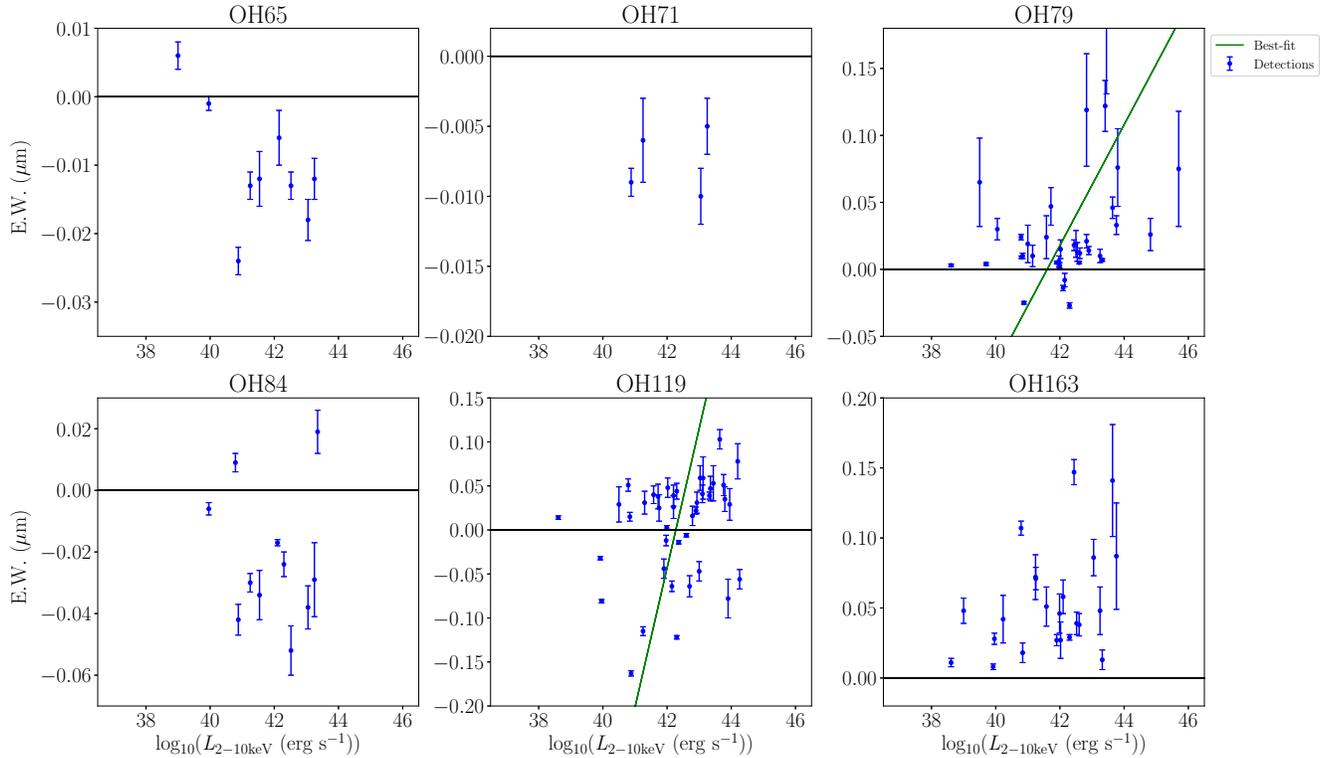}
    \caption{EW of the OH doublet vs. log($L_{\rm{2-10keV}}$ (erg/s)).  Galaxies must have detections in EW(OH) and $L_{\rm{2-10keV}}$ to be included in the plot and fits.  The weighted log-linear fits, using the parameters given in Table \ref{tab:luminosity_stats}, are shown for fits with p $<$ 0.05.  For both EW(OH79) and EW(OH119), $|r|$ $\leq$ 0.5, which indicates that $L_{\rm{2-10keV}}$ is an uncertain predictor of these doublets.  We do not show the fit for slopes with a p-value $>$ 0.05.  For OH79, MCG-01-24-012 is an outlier (EW = 0.242$\mu$m), and is cut off to better show the rest of the data.}
    \label{fig:L_2_10_plot}
\end{figure*}

\begin{figure*}[t!]
    \includegraphics[scale=0.36]{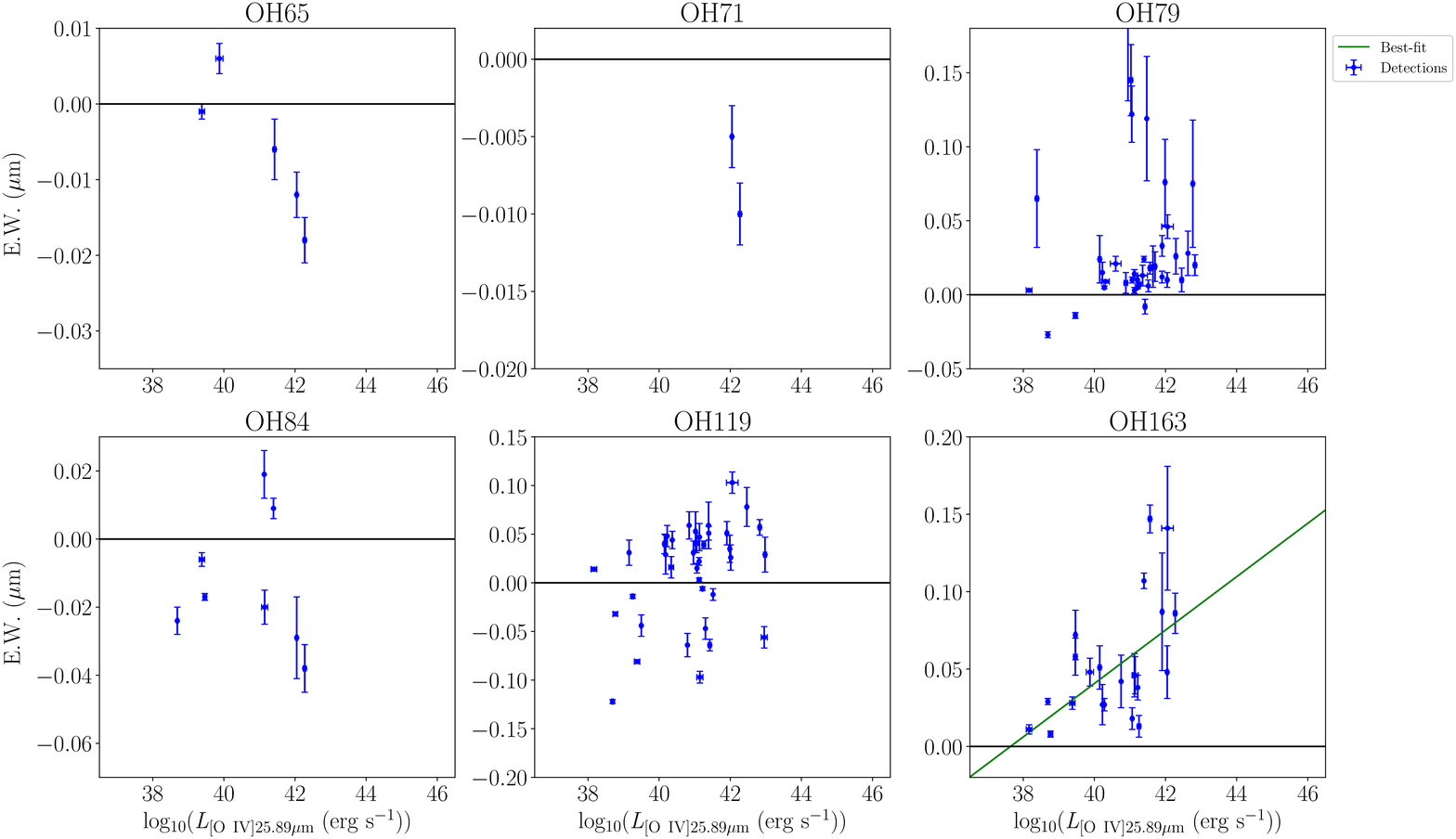}
    \caption{EW of the OH doublet vs. log($L_{\rm{[O\ IV]25.89\mu m}}$ (erg/s)) gathered from IRAS.  Galaxies must have detections in EW(OH) and $L_{\rm{[O\ IV]25.89\mu m}}$ to be included in the plot and fits.  The weighted log-linear fits, using the parameters given in Table \ref{tab:luminosity_stats}, are shown for fits with p $<$ 0.05.  We do not show the fit for slopes with a p-value $>$ 0.05.  For OH79, MCG-01-24-012 is an outlier (EW = 0.242$\mu$m), and is cut off to better show the rest of the data.}
    \label{fig:L_IR_plot}
\end{figure*}

\subsubsection{AGN Luminosity} \label{subsec:agn_luminosity}

To probe the brightness of the AGN, we convert the most highly ionized line fluxes from Section \ref{subsec:line_ratios}, [O~\textsc{IV}] 25.89$\mu$m, [Ne~\textsc{V}] 14.32$\mu$m, and [Ne~\textsc{V}] 24.32$\mu$m, to luminosities.  As outlined in Section \ref{subsec:luminosities_and_IR_line_ratios}, we will refer to these as $L_{\rm{[O\ IV]25.89\mu m}}$, $L_{\rm{[Ne\ V]14.32\mu m}}$, and $L_{\rm{[Ne\ V]24.32\mu m}}$, respectively.  We will also investigate possible correlations with the intrinsic hard X-ray (2-10 keV) luminosity, $L_{\rm{2-10keV}}$.  These luminosities should all increase with increasing AGN activity in the host galaxy (e.g. \citealt{tom08, tom10}; also \citealt{dia09} for $L_{\rm{[O\ IV]25.89\mu m}}$).  

We use the same fitting method as Section \ref{subsec:line_ratios}.  Unlike the IR luminosities, the $L_{\rm{2-10keV}}$ data was published without estimated uncertainties.  Therefore, in this scenario, only the uncertainty on EW(OH) is incorporated into the fits.  Table \ref{tab:luminosity_stats} in the Appendix gives the results for the log-linear fits.  We include plots for EW(OH) vs. log($L_{\rm{2-10keV}}$) (Figure \ref{fig:L_2_10_plot}) and log($L_{\rm{[O\ IV]25.89\mu m}}$) (Figure \ref{fig:L_IR_plot}), with one panel for each OH line in this section.  Figures for EW(OH) vs. the other two line ratios ($L_{\rm{[Ne\ V]14.32\mu m}}$ and $L_{\rm{[Ne\ V]24.32\mu m}}$) are available as supplementary material in the online version of the paper.  The fits from Table \ref{tab:luminosity_stats} with p $<$ 0.05 are shown in the Figures as a solid green line.

We find very few correlations between the OH doublets and the IR and X-ray luminosities.  EW(OH163) displays a significant correlation ($r$ = 0.56, p = 0.007) with log($L_{\rm{[O\ IV]25.89\mu m}}$).  This is the strongest correlation found between an OH doublet and a luminosity observable.  Due to low p-values (0.03 and 0.014) but large scatter in the data ($r$ = 0.35 and 0.41), log($L_{\rm{2-10keV}}$) is an uncertain predictor of EW(OH119) and EW(OH79), respectively.  Despite the scatter for OH79 and OH119, these relationships indicate that these three OH doublets all display the strong emission in the most luminous AGN.  At lower luminosities, which indicate either weak or no AGN activity, we see deep absorption (OH119), weak emission (OH163), and a mixture of weak emission and weak absorption (OH79).

Similar to Section \ref{subsec:line_ratios}, no correlations are found for OH65, OH71, and OH84.  Low statistics once again are a factor in this.  It is possible that a relationship between EW(OH65) and log($L_{\rm{[O\ IV]25.89\mu m}}$) could exist ($r$ = $-$0.90, p = 0.04), but this is based on only five galaxies.  However, if this correlation exists, it suggests that a luminous source would be needed to create gas temperature warm enough to drive OH65 into absorption.

\subsubsection{Comparison Between the Luminosity and IR Emission-Line Tracers} \label{subsec:luminosity_line_ratio_comparison}

The IR emission-line ratios in Section \ref{subsec:line_ratios} and luminosities in Section \ref{subsec:agn_luminosity} both trace the strength of the AGN to probe the spectral type of the galaxy.  When OH79, OH119, and OH163 correlate with one of these AGN tracers (luminosity or emission-line ratio), it is always positive.  This means that these doublets display the strongest emission when the emission-line ratios and luminosities suggest that the galaxy contains a dominant AGN feature.  When the tracers suggest that the galaxy has either a weak or no AGN, we observe deep absorption (OH119), weak emission (OH163), and a mixture of weak emission and weak absorption (OH79).  The slopes of the fits are generally largest for EW(OH119) meaning that this doublet is the most affected by the AGN tracers.  Because we observe EW(OH119) with deep absorption and strong emission, and do not for EW(OH79) and EW(OH163), this is reasonable.  For OH119, the results that emission is stronger in more luminous AGN aligns with previous studies \citep{vei13, sto16}.

OH79, OH119, and OH163 all correlate with a higher percentage of the emission line ratios than the X-ray and IR luminosities.  Also, the relationships with the IR emission-line flux ratios are generally at a higher significance with less scatter.  These two results suggest that comparing the relative contributions of AGN activity and star formation is a better predictor of EW(OH) than the AGN luminosity.  In other words, EW(OH) cares more about the relative contribution of the AGN to the total power (the line ratios), and less about the total power of the AGN (the luminosities).  This result is in line with \citet{flu19}, who find that the mass loading factor (which measures how fast the outflow ejects the gas from the galaxy compared to the gas consumption by star formation) depends on the relative contribution of the AGN to the bolometric luminosity.

While not significant due to samples of only five galaxies, EW(OH65) is also found to anti-correlate with both IR emission-line ratio and IR luminosity tracers.  As discussed above, this suggests that an AGN is needed to elevate the temperature of gas high enough ($\sim$300 K), to drive OH65 into absorption.

\subsection{EW(OH) vs. 25-60 Color Temperature} \label{subsec:flux_density}

\begin{figure*}[t!]
    \centering
    \includegraphics[scale=0.36]{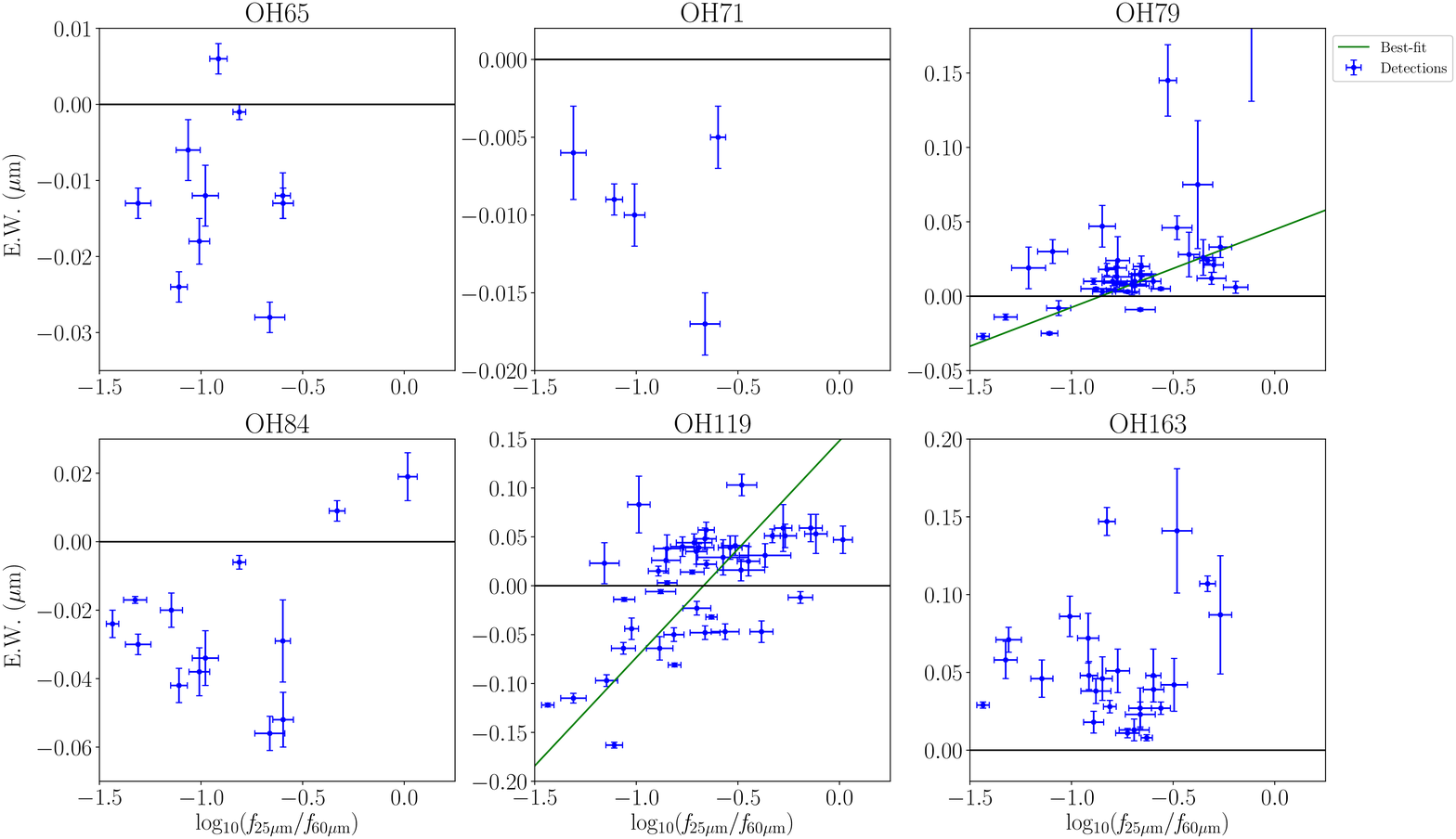}
    \caption{EW of the OH doublet vs. the logarithm of the dust temperature traced by the IRAS $f_{25\mu\rm{m}}$/$f_{60\mu\rm{m}}$ ratio.  Galaxies must have detections in EW(OH), IRAS 25$\mu$m flux, and IRAS 60$\mu$m flux to be included in the plot and fits.  The weighted linear fits, using the parameters given in Table \ref{tab:flux_stats}, are shown for fits with p $<$ 0.05.  We do not show the fit for slopes with a p-value $>$ 0.05.  For OH79, MCG-01-24-012 is an outlier (EW = 0.242$\mu$m), and is cut off to better show the rest of the data.}
    \label{fig:iras_plot}
\end{figure*}

In this section, we estimate the thermal dust continuum emission to use as another IR probe of AGN strength.  As stated in Section \ref{subsec:IRAS}, to investigate the dominance of the AGN component in the host galaxy, we use the $f_{25\mu\rm{m}}$/$f_{60\mu\rm{m}}$ ratio (referred to as \textquotedblleft dust temperature") from IRAS.  The ratio of these flux densities can be used as a proxy of the strength of the AGN because the majority of the 25$\mu$m flux originates from warm dust surrounding the AGN while most of the 60$\mu$m flux is from cooler dust in star formation.  Therefore, we expect the galaxies with luminous AGN to have a higher dust temperature than less luminous AGN and non-active galaxies (e.g. \citealt{ede86, spi95, vei09, lee11}).

We fit the data using the same method described in the previous section, using uncertainties for both EW(OH) and dust temperature.  Table \ref{tab:flux_stats} provides the parameters of the best-fit lines and Spearman rank-order correlation statistics, and a plot of EW(OH) vs. the logarithm of the dust temperature traced by the IRAS $f_{25\mu\rm{m}}$/$f_{60\mu\rm{m}}$ ratio is shown in Figure \ref{fig:iras_plot}.   

We find a highly significant ($r$ $\geq$ 0.5, p $<$ 0.01) positive relationship between dust temperature and both EW(OH79) and EW(OH119).  These correlations show that the two OH doublets are more likely to be in strong emission when surrounded by warmer dust, which indicates that the AGN component has a relatively high luminosity output.  For galaxies with cooler dust, indicating a galaxy with either a weak or no AGN component, we find OH119 primarily in deep absorption.  OH79 is shown to be in either weak absorption, compared to OH119, or weak emission when surrounded by cooler dust temperatures.

The other four OH doublets show no correlation with dust temperature.  For OH65, \citet{gon15} did not find a significant relationship between EW and dust temperature as well.  The samples are similar in size: there are 23 galaxies (10 detections) in this sample and 29 (25 detections) in \citet{gon15}.  This study contains 9 galaxies ($\sim$39\%) not included in \citet{gon15}; however, 8/10 detections in this study are also in \citet{gon15}.

\subsection{EW vs. Dust Obscuration} \label{subsec:red_test}

This section investigates possible effects of dust reddening and extinction on the sample using one IR and one optical dust tracer.

\subsubsection{IR Silicate Absorption} \label{subsec:silicates}

\begin{figure*}[t!]
    \centering
    \includegraphics[scale=0.36]{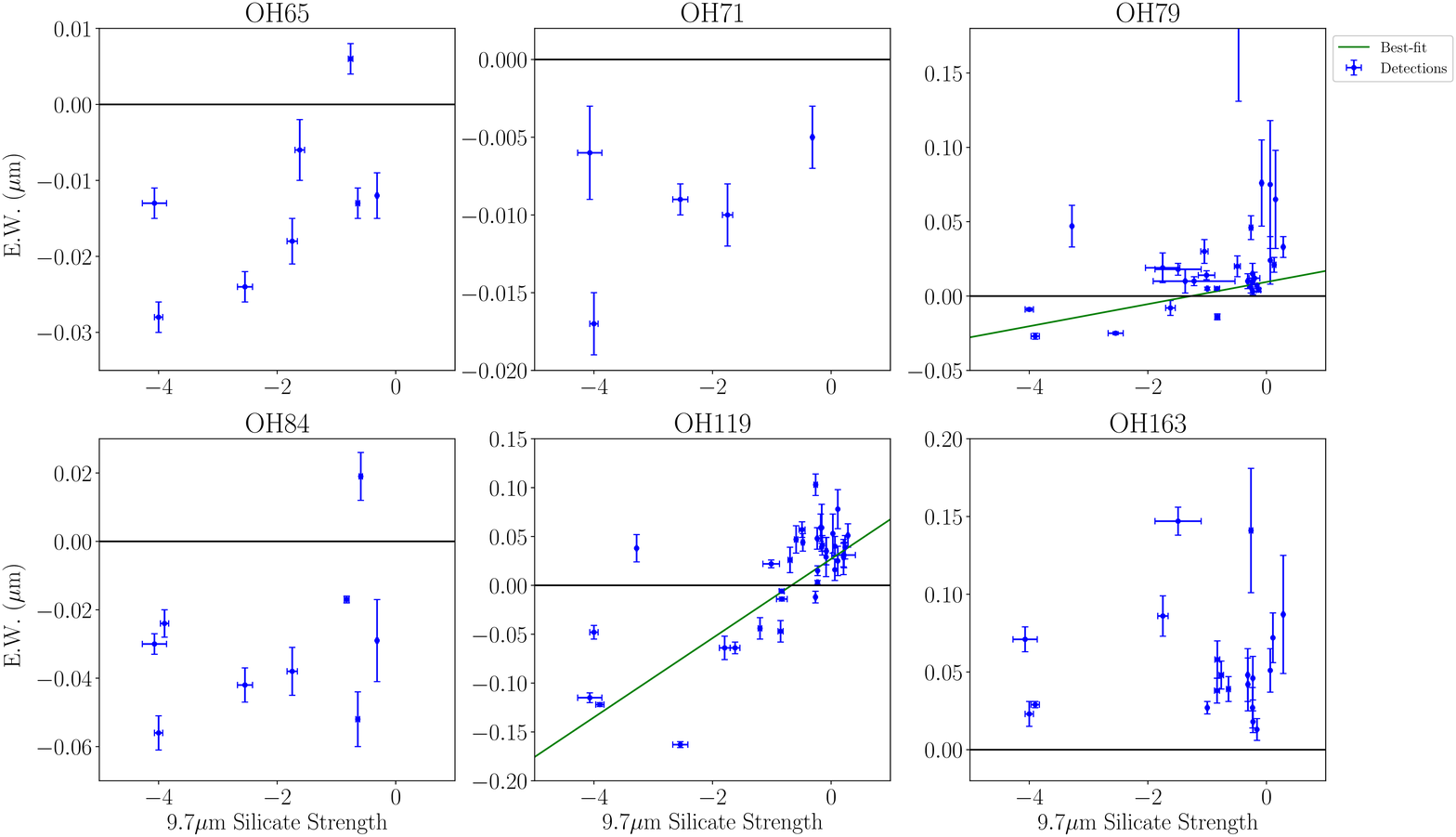}
    \caption{EW of the OH doublet vs. the strength of the 9.7$\mu$m silicate feature.  Galaxies must have detections in EW(OH) and $S_{\rm{sil}}$ to be included in the plot and fits.  The weighted log-linear fits, using the parameters given in Table \ref{tab:extinction_stats}, are shown for fits with p $<$ 0.05.  We do not show the fit for slopes with a p-value $>$ 0.05.  For OH79, MCG-01-24-012 is an outlier (EW = 0.242$\mu$m), and is cut off to better show the rest of the data.}
    \label{fig:silicate_plot}
\end{figure*}

The OH doublets we observe are produced by warm/dense molecular gas, which is expected to be associated with dust.  We therefore, compared our OH measurements with the 9.7$\mu$m silicate dust feature in those same galaxies.  Recall from Section \ref{subsec:silicates_and_balmer_data} that the measurements were compiled from a variety of sources (\citealt{alo16, far13, gm15, her15, wu09}) (see Table \ref{tab:silicate_and_balmer_data}) and the strength of the silicate feature is quantified using the method derived by \citet{spo07}.  Figure \ref{fig:silicate_plot} shows the EW plotted against the strength of the silicate feature, and Table \ref{tab:extinction_stats} contains the fitting parameters and correlation statistics to the data.

We find a highly significant correlation ($r$ = 0.59, p = 0.001) for EW(OH119).  The relationship is positive, meaning that as $S_{\rm{sil}}$ increases (silicate absorption weakens), this doublet is more likely to display stronger emission.  It is important to note that this relationship can be strengthened by removing one outlier, IRAS 11095-0238.  We identify this galaxy as an outlier because OH119 is in emission even though it has the fourth strongest silicate absorption in the sample ($S_{\rm{sil}}$ = $-3.28$).  Except for Mrk 1066 ($S_{\rm{sil}}$ = $-1.01$) no other galaxy shows EW(OH119) past a silicate strength of $-1.00$.  Excluding IRAS 11095-0238, increases the strength of the fit: $r$ = 0.65, p = 0.00003.

The results for OH119 agree with \citet{sto16} (which expand on/confirm the results of \citealt{spo13} and \citealt{vei13}).  For OH119, these studies find that a stronger silicate absorption feature (more dust extinction), leads to weaker OH119 emission and stronger OH119 absorption.  \citet{spo13} and \citet{vei13} find this correlation using samples of local ULIRGs.  \citet{sto16} combines the sample of ULIRGs from \citet{vei13} and incorporates their BAT AGNs to create a more expansive sample.   Our sample size is comparable in size to \citet{sto16}, with some overlap in the selected galaxies, allowing us to confirm this relationship.  

We find $S_{\rm{sil}}$ to be a uncertain predictor ($r$ = 0.42, p = 0.02) of EW(OH79).  However, removing the two largest outliers, IRAS 11095-0238 (similar reasons as for OH119) and MCG-01-24-012 (due to extreme EW emission, +0.242), increases the strength of the relationship, and we now find a significant correlation ($r$ = 0.54, p = 0.002) between the two observables.  Similar to OH119, the relationship between EW(OH79) and $S_{\rm{sil}}$ is positive; however, the $r$ value for OH79 is lower than that of OH119.  This comparison exists both with and without the outlier galaxies in the two samples, meaning that the strength of the 9.7$\mu$m silicate feature is more predictive of EW(OH119).

The other four OH doublets show no correlation with $S_{\rm{sil}}$, based on their p-values.  Our results for EW(OH65) do show a high correlation coefficient ($r$ = 0.60); however, this does not translate to a small p-value (p = 0.12) because there are only eight galaxies in the sample.  \citet{gon15} had almost three times the number of detections as this study, with six of our eight detections in their sample, and they do find a correlation between EW(OH65) and silicate strength.  They claim that while deep OH65 absorption does not guarantee deep silicate absorption, a trend between the two exists.  \citet{gon15} also claim that the two quantities are produced from the same material.  Our sample does not contradict this, but is too small to be conclusive.

\subsubsection{Balmer Decrement} \label{subsec:balmer_extinction}

An additional optical tracer for the extinction affecting the ionized gas is the H$\alpha$/H$\beta$ ratio.  H$\beta$ is more obscured by dust, so it is expected that a larger Balmer ratio will be observed in dustier galaxies.  The line fluxes were collected from published measurements in the literature, and Table \ref{tab:silicate_and_balmer_data} provides the data measurements along with the citations.  

The results of the linear regression analysis are given in Table \ref{tab:extinction_stats} in the Appendix, and a figure for EW(OH) vs. H$\alpha$/H$\beta$ is available as supplementary material in the online version of the paper.  EW(OH79) has the strongest correlation with the Balmer decrement in the sample.  The correlation is significant ($r$ = $-$0.63, p = 0.002), and reveals that a larger H$\alpha$/H$\beta$ ratio (i.e. more dust obscuration) results in either weak emission or weak absorption.  The galaxies with the strongest EW(OH79) emission have small Balmer ratios.

Similar to OH79, OH119 tends to have the strongest emission at low H$\alpha$/H$\beta$, and emission gets weaker and absorption becomes more prevalent at high H$\alpha$/H$\beta$.  The relationship between EW(OH119) and H$\alpha$/H$\beta$ has a low p-value (0.014) but also a low $r$ value ($-$0.47), which indicates that the Balmer decrement is an uncertain predictor of the doublet.  However, the correlation greatly improves when removing two outliers, NGC 253 and NGC 7172.  These two galaxies can be identified as outliers in the sample because they have deep absorption ($-$0.081$\mu$m and $-$0.064$\mu$m) at low H$\alpha$/H$\beta$ ratios (2.84 and 3.00).  All other galaxies in the sample show strong emission (EW $\sim$ +0.05$\mu$m) in this Balmer decrement range.  Without NGC 253 and NGC 7172, EW(OH119) has a significant relationship ($r$ = $-$0.72 and p = 0.00005) with the Balmer decrement.  NGC 7172 was identified in Section \ref{subsec:line_ratios} as a galaxy with a recent AGN source that needs some Myrs to drive OH119 into emission.  This means that even though this galaxy has low dust extinction, the nuclear source was not powerful enough to create OH119 emission until just recently and the doublet needs more time to reach equilibrium in emission.

While both the silicate feature and Balmer decrement show that a higher dust concentration leads to weaker emission and deeper absorption for the OH79 and OH119 doublets, it is shown that $S_{\rm{sil}}$ is the better extinction predictor for EW(OH119) and the Balmer decrement is the better predictor for EW(OH79).

No correlation is found for the other four OH transitions in the sample.  For these doublets, OH163 was the only one to meet the eight galaxies threshold; however, even with small statistics, the other three galaxies show no indication of a possible relationship with the Balmer decrement.

\subsection{Bivariate Linear Regression Test for EW(OH)} \label{subsec:bivariate}

In Sections \ref{subsec:line_ratios}, \ref{subsec:agn_luminosity}, and \ref{subsec:red_test}, we found correlations between the EW of OH doublets and: 1) IR and X-ray luminosities of the galaxy; 2) IR emission-line ratios; and 3) dust extinction tracers.  In those sections we found many correlations of varying strength between EW(OH) and the many observables.  Here we use a bivariate linear regression to see if EW(OH) is correlated with multiple observables.  We chose one observable from each section: 1) the $L_{\rm{2-10keV}}$ to probe the brightness of the AGN using X-rays; 2) the log([O~\textsc{IV}] 25.89$\mu$m/[Ne~\textsc{II}] 12.81$\mu$m) IR emission-line ratio because this is a tracer of the relative AGN/star-formation contribution; and 3) the strength of the silicate feature for the dust extinction tracer because this traces the obscuration.  The dependent variable in the bivariate linear regression, EW(OH), is weighted using its uncertainties.  Due to limited sample size, we do not include OH65, OH71, and OH84 in these bivariate relationships.  Table \ref{tab:biv_table} gives the coefficient of each independent parameter ($L_{\rm{2-10keV}}$, log([O~\textsc{IV}] 25.89$\mu$m/[Ne~\textsc{II}] 12.81$\mu$m), and $S_{\rm{sil}}$), the y-intercept of the regression, and the $R^2$ value evaluating the strength of the correlation.

There is a very strong correlation between EW(OH119) and the bivariate linear regression ($R^2$ = 0.91).  This makes sense because this doublet has a relationship with all three observables individually.  While not as strong as the OH119 doublet, EW(OH79) is correlated with this bivariate linear regression ($R^2$ = 0.61).  OH79 also displayed a relationship with all three observables individually; however, these were generally not as strong as those for EW(OH119).  This could explain why the bivariate correlation for EW(OH79) is weaker.  The bivariate regression shows no correlation with EW(OH163), which can be expected since this doublet displayed no correlation with any of the three observables individually.

\subsection{OH79 vs. OH119} \label{subsec:79_vs_119}

\begin{figure}[ht!]
    \centering
    \includegraphics[width=0.95\linewidth]{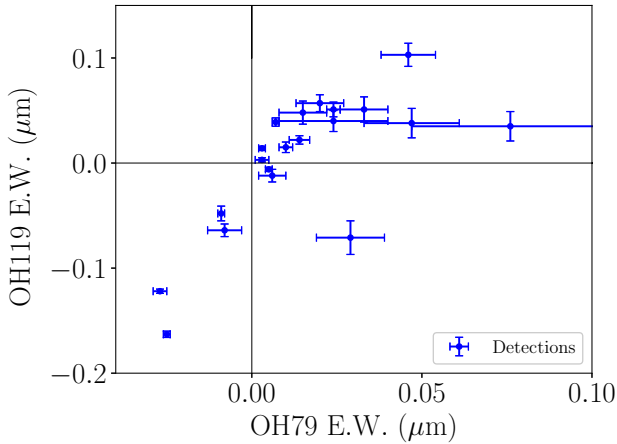}
    \caption{EW(OH119) vs. EW(OH79).  Only galaxies with detections for both doublets are shown.  The error bars for Mrk 478 is cut off to better show the rest of the data.}
    \label{fig:79_vs_119_plot}
\end{figure}

The OH79 and OH119 doublets show the most significant correlations in Sections \ref{subsec:width_vs_type}, \ref{subsec:flux_density}, and \ref{subsec:red_test}.  Also, the transitions share the same lower energy level, suggesting a tight relationship between both OH doublets.  To test this, we compared the EW for the 20 galaxies that have detections for both transitions (this excludes galaxies with P-Cygni features and upper limit observations).  Figure \ref{fig:79_vs_119_plot} displays the plot comparing the EW of the OH79 and OH119.

Figure \ref{fig:79_vs_119_plot} reveals that for most galaxies, the doublets are either both in emission or both in absorption, and a larger EW for one transition generally indicates a larger EW for the other.  This positive trend between the two doublets is very strong with high significance, as indicated by $r$ = 0.71 and p = 0.0005.  There does seem to be a saturation at EW(OH79) $\gtrsim$ 0.03$\mu$m, where OH119 seems to flatten around EW(OH119) $\sim$ 0.05$\mu$m.  In the sample, there are three objects that have an emission profile for OH79 and an absorption profile for OH119: IRAS 20037-1547, PKS 2048-57, and NGC 7582.  Two of them, PKS 2048-57 and NGC 7582, have small EW(OH79) emission, small EW(OH119) absorption, and appear to follow the trend with the rest of the sample.  The third galaxy, IRAS 20037-1547, however, is a noticeable outlier.  There are no galaxies that have an absorption profile for OH79 and an emission profile for OH119.

\subsection{OH79 vs. OH163}

\begin{figure}[ht!]
    \centering
    \includegraphics[width=0.95\linewidth]{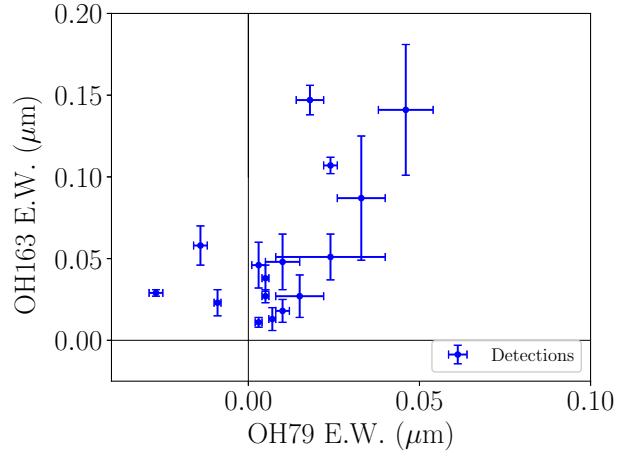}
    \caption{EW(OH163) vs. EW(OH79).  Only galaxies with detections for both doublets are shown.}
    \label{fig:79_vs_163_plot}
\end{figure}

OH79 and OH163 are the two doublets most likely to be seen in emission in the sample.  They also are connected in the OH Grotrian diagram, as the upper level of the OH79 doublet is the lower level of the OH163 doublet.  To test if there is a relationship between EW(OH79) and EW(OH163), we look at the 16 galaxies (excluding galaxies with P-Cygni features and upper limit observations) that have detections for both transitions.  Figure \ref{fig:79_vs_163_plot} shows the plot comparing the EWs.  

Figure \ref{fig:79_vs_163_plot} shows that there is weak correlation ($r$ = 0.54, p = 0.03) between EW(OH79) and EW(OH163).  The trend is positive, meaning that stronger emission for one of the doublets implies stronger emission for the other.  This relationship could explain why we only observe OH163 in emission in the complete sample.  The lower level of the OH163 doublet could be depopulated by the OH79 transition, which would prevent OH163 absorption from occurring.  

There are three objects where OH79 shows absorption while OH163 shows emission: NGC 3079, NGC 4418, and NGC 4945.  All three of these galaxies have buried nuclei, indicated by their deep 9.7$\mu$m silicate absorption.  NGC 4418 and NGC 4945 have the second and third strongest silicate absorption in the entire 178 galaxy sample.  Removing these three galaxies reveals a significant correlation ($r$ = 0.73, p = 0.004) between the OH163 and OH79 doublets.

\section{Discussion and Interpretations} \label{sec:discussion}

To gain a deeper understanding of the correlations that we found in Section \ref{sec:analysis_and_results}, we look to compare our results with various models of OH transitions from the literature.  We can interpret the results described above in terms of OH level populations because we combine several transitions measured simultaneously.  

\subsection{OH Level Populations \& the Mechanisms which Control them} \label{subsec:RT_models}

OH emission can arise from two mechanisms: radiative pumping and collisional excitation.  Collisions with the surrounding medium are dependent on the temperature of the region.  The nuclear region of an active galaxy is warmer than that of a star-forming galaxy, therefore the former will have more energetic collisions.  The correlations of EW(OH) with the galaxy properties that we analyzed in Section \ref{sec:analysis_and_results} are probably driven by the nuclear region of the galaxies, where the impact of the AGN is dominant.

To estimate the energy of possible collisions, we reference the results of models from \citet{gon04, gon08, gon12, gon14, gon17a} and \citet{fal15} that generate the observed OH transitions to learn about properties of the environment that they were created in.  The models in these studies have multiple components; they start with a hot inner region and cooler extended region, and then add one or two extra components if necessary to better fit the observed spectra.  One example of an extra component added to the two-region model is the addition of a more extended halo component with a lower temperature to Arp220 \citep{gon04}.  For most of the galaxies modeled in these studies, the dust temperature of the far-IR continuum source ranges from about 90$-$150 K for the inner nuclear region, while the cooler extended region is roughly 40$-$90 K.

For the OH transitions in this sample, radiative excitation primarily comes from far-IR photons, which can originate from either AGN activity or star formation.  We will discuss the method of excitation for the OH transitions in our sample one at a time.  We refer to the OH Grotrian diagram, Figure \ref{fig:energy_diagram} in Section \ref{subsec:PACS}, throughout this section.

\subsubsection{OH119}

Sections \ref{subsec:line_ratios}, \ref{subsec:agn_luminosity}, and \ref{subsec:flux_density} probe the AGN luminosity, radiation field hardness, and dust temperature of a galaxy.  The results in these Sections show that EW(OH119) has a significant correlation ($r$ $\geq$ 0.5, p $<$ 0.01) with 3/5 IR emission-line ratios and the IRAS $f_{25\mu\rm{m}}$/$f_{60\mu\rm{m}}$ dust temperature.  An additional IR line ratio and the intrinsic 2-10 keV X-ray luminosity are also uncertain predictors of EW(OH119).  Therefore, 6/10 observables tracing the strength of an AGN have some level of predictive power for EW(OH119).  The relationships are positive, meaning that when these observables are large (luminous AGN, hard radiation field, and warm dust), OH119 is very likely to be in emission, and the opposite (dim/no AGN, soft radiation field, cooler dust), tends to bring OH119 into deep absorption.  These results indicate that OH119 emission is dependent on the conditions created by an AGN, which suggests that OH119 emission is created through collisional excitation.  If OH119 emission were generated primarily by radiative pumping of far-IR photons, we would expect to find the doublet in emission for galaxies with a weak or no AGN, which we do not find.  However, a possible contribution to the OH119 emission from radiative pumping by far-IR photons cannot be disregarded.

This idea agrees with \citet{spi05} and \citet{sto16}, which both claim that OH119 emission is generated through collisions.  The argument from \citet{spi05} comes from the relative strengths of the OH transitions they observed and the Einstein-A coefficients.  Table \ref{tab:oh_lines} in Section \ref{sec:intro} gives the Einstein-A coefficients of the OH transitions in this study, which were gathered from the LAMBDA database \citep{sch05}.  As discussed in \citet{spi05}, if the OH119 transitions were excited exclusively via radiative pumping, there would be two paths it could take: absorption of either the 34.6$\mu$m or 53.3$\mu$m ground-state OH transitions.  If the radiative pumping happened through the 34.6$\mu$m transition, then the 98.7$\mu$m OH transition would be about five times stronger than OH119.  If OH119 were excited by the 53.3$\mu$m path, then OH163 would be about five times stronger than OH119.  The estimate of the relative strengths between the OH transitions come from the Einstein-A coefficients.  In their study, \citet{spi05} could not detect the 98.7$\mu$m doublet, and OH163 was not five times stronger than OH119, so they concluded that OH119 must be a result of collisional excitation, and not radiative pumping.

Our results agree with the claims made by \citet{spi05} on how the OH119 comes into emission.  When comparing galaxies that have detections showing emission for both OH119 and OH163, we find median fluxes of 4.44 $\times$ 10$^{-14}$ erg s$^{-1}$ cm$^{-2}$ and 2.62 $\times$ 10$^{-14}$ erg s$^{-1}$ cm$^{-2}$ for the sample.  This shows that the OH119 doublet is actually stronger on average than the OH163 doublet when comparing the emission detections in the same galaxy (the average OH163/OH119 flux ratio is 0.59).  The only galaxy where the OH163 doublet is stronger is NGC 1365.  Because we do not see a flux ratio of $\sim$5 for OH163/OH119, it seems unlikely that the 53.3$\mu$m path dominates the OH119 emission via radiative pumping.  Unfortunately, we do not have any OH 98.7$\mu$m detections, similar to \citet{spi05}, to confirm or rule out the 34.6$\mu$m path as a possible source of OH119 emission from radiative pumping; however, based on the results with our AGN tracers radiative pumping through the 34.6$\mu$m path seems unlikely.

In \citet{sto16}, they use their correlation between EW(OH119) and the silicate feature to conclude that OH119 emission comes from a circumnuclear region where the temperature, OH abundance, and number density favors collisional excitation over radiative pumping.  We find a similar relationship between EW(OH119) and the silicate feature, and therefore also support the idea that OH119 emission originates from collisions.

These correlations, which indicate the importance of collisional excitation of OH119, are nonetheless imperfect, and some of the observed scatter could be attributed to other factors.  In particular, high obscuration is correlated with absorption in this transition.  Therefore, where that additional affect is strong, i.e. in a heavily obscured AGN, the OH doublet may not be strongly in emission.

\subsubsection{OH163}

As stated in Section \ref{subsec:oh163}, all detections for the OH163 doublet are in emission.  Because the sample spans all Seyfert types and includes LINERs and non-active galaxies, this means that an AGN is not needed to generate OH163 emission.  In fact, we do not find a correlation between EW(OH163) and the majority of the AGN and dust temperature tracers discussed in Section \ref{sec:analysis_and_results}.  Therefore, OH163 emission should arise primarily from radiative pumping of the far-IR background, which as discussed above can come from either AGN activity or star formation.

Once again, this idea agrees with \citet{spi05}, who also claim that OH163 emission arises primarily from radiative pumping.  For this doublet, the upper level of the transition is 270 K above the ground state.  \citet{spi05} argue is too high for collisional excitation to be effective; however, cascading down from absorption of a 34.6$\mu$m or 53.3$\mu$m photon can lead to OH163 emission.  This is a reasonable assumption given the ranges of $T_{dust}$ in the models referenced at the beginning of this section.

The correlations between EW(OH163) and only a few of the AGN tracers $-$ log([Ne~\textsc{V}] 24.32$\mu$m/[Ne~\textsc{II}] 12.81$\mu$m), log([Ne~\textsc{III}] 15.56$\mu$m/[Ne~\textsc{II}] 12.81$\mu$m), and log($L_{\rm{[O\ IV]25.89\mu m}}$) $-$ could suggest that the inclusion of an AGN creating higher luminosities could lead to more far-IR photons available to pump OH163 into stronger emission.  Without an AGN, there will still be far-IR photons for radiative pumping $-$ albeit there will be fewer $-$ which is why we see OH163 in weaker emission, but still emission nonetheless.  No correlation is found between EW(OH163) and the majority of the AGN tracers however, which could be due to the fact that star formation can also affect the strength of the far-IR background.

\subsubsection{OH79}

The results in Sections \ref{subsec:line_ratios}, \ref{subsec:agn_luminosity}, and \ref{subsec:flux_density} show that EW(OH79) has a significant correlation with one of the IR emission-line ratios, [Ne~\textsc{V}] 14.32$\mu$m/[Ne~\textsc{II}] 12.81$\mu$m, and the IRAS $f_{25\mu\rm{m}}$/$f_{60\mu\rm{m}}$ dust temperature.  Two more IR line ratios and the intrinsic 2-10 keV X-ray luminosity are also uncertain predictors of the doublet.  These relationships are all positive, so EW(OH79) emission is strongest when these observables are large (luminous AGN with warm surrounding dust), similar to OH119 and OH163.  When we observe the opposite (dim/no AGN with cooler surrounding dust) some galaxies show OH79 in weak emission (similar to OH163) and other galaxies show OH79 in weak absorption (compared to OH119).  One possible explanation for this difference could be that emission for the OH79 doublet arises from a mix of collisional excitation and radiative pumping.  The effects of the latter could prevent OH79 from being observed in deep absorption when an AGN is not present (similar to OH119), and instead keeping the doublet in either weak emission or weak absorption.

This is supported when comparing EW(OH79) and EW(OH119) in Section \ref{subsec:79_vs_119}.  In this smaller subsample of galaxies that have detections of both doublets, there are seven galaxies where OH119 is in absorption, and OH79 is in absorption in four of them.  Also, while the peak emission for both doublets is similar ($\sim$0.10), but the deepest absorption is about an order of magnitude different ($\sim -$0.15 for EW(OH119) and $\sim -$0.03 for EW(OH79)).  This suggests that while the presence of a dominant AGN affects both doublets similarly, radiative pumping from a far-IR background can keep OH79 in weak emission when there is either a weak or no AGN present.

This argument that OH79 emission arises from a mix of radiative pumping and collisions was first made by \citet{spi05}.  The upper level of the OH79 transition is only 182 K above the ground state, which \citet{spi05} argue is a low enough temperature for a warm dense region (AGN) to excite it, at least partially, through collisional excitation.  Also, OH79 could be excited through the same far-IR radiative pumping mechanism as OH163 (i.e. cascading down from absorption of a 34.6$\mu$m or 53.3$\mu$m photon).

\subsubsection{OH65}

For OH65, we make an argument based on energy levels to rule out collisions as the primary form of excitation.  The temperatures derived from the literature models are much lower than what would be needed to collisionally excite OH65 (the upper level of the transition is $\sim$500 K) in any significant amount.  While AGNs cannot create warm enough environments to generate OH65 emission, the results from correlating EW(OH65) with log([O~\textsc{IV}] 25.89$\mu$m/[Ne~\textsc{II}] 12.81$\mu$m), log([O~\textsc{IV}] 25.89$\mu$m/[O~\textsc{III}] 88.36$\mu$m) and log($L_{\rm{[O\ IV]25.89\mu m}}$) suggest that AGNs could drive OH65 absorption.  The sample sizes for these correlations are very small (only 5 galaxies); however, there is a strong ($r$ = $-$0.90, p = 0.04) negative relationship between EW(OH65) and each of these observables.  These results suggest that a warm AGN region is needed to populate the lower energy level ($\sim$300 K) of the OH65 doublet.  In a weak radiation field (i.e. a non-active galaxy), the population of OH molecules in the lower level of the 65$\mu$m transition could be scarce, thus suppressing the absorption in this doublet.  The one galaxy in the sample with OH65 in emission is NGC 2146, which is a starburst galaxy, supports this idea.  Based on the temperatures from the literature, the emission in NGC 2146 is likely not a result of collisional excitation.  Radiative pumping is therefore the more likely method.

\subsubsection{OH71}

OH71 is the highest energy transition in the sample, with the upper level of the transition $\sim$600 K.  Therefore, this temperature is too high for significant amounts of collisionally excited OH71 emission, similar to OH65.  It is possible that we only observe OH71 exclusively in absorption because it is such a high energy transition.  OH71 has the smallest number of detections in the sample, so it is possible that more observations could reveal rare cases of emission.  When investigating possible correlations with AGN tracers, the statistics were too small (three or fewer galaxies) for the OH71 doublet to investigate if a similar phenomenon to OH65 is observed.  More data are needed to see if AGNs drive OH71 absorption, similar to OH65.

\subsubsection{OH84}

For OH84, it is difficult to strongly support either collisional excitation or radiative pumping as the main mode of excitation.  On the energy diagram, OH84 does not lie significantly above the temperature that the nuclear region is thought to have, so at least a small amount of collisional excitation could be possible in the warmest galaxies.  We did not find a relationship with any of the AGN, radiation field hardness, and dust temperature tracers; however, small sample sizes for many of these test make it difficult to draw conclusions.  More data are needed to make a strong argument about whether OH84 emission arises from collisional excitation or radiative pumping.

\subsection{Interpretation of the Obscuration \& Column density} \label{subsec:obscuration_takeaways}

A high column density of gas and dust might be expected to enhance both the absorption strength of OH and the dust obscuration.  We used two observables as proxies for the column density of interstellar matter along our line-of-sight to the center of the galaxy.  The H$\alpha$/H$\beta$ ratio measures dust extinction since it increases from its intrinsic ratio (which may range from $\sim$3 in low-density ionized gas to values greater than 4 in the BLR of a Seyfert 1 nucleus \citep{lac82, mal17}.  The brighter the AGN, the more concentrated this line emission should be in the nucleus.  For extremely dusty galaxies (e.g. ULIRGs), the optical line ratios could substantially under-estimate the dust column density to the galaxy center.  

At high extinctions, a reliable tracer in the mid-IR range is the depth of the silicate absorption at 9.7$\mu$m.  In Seyfert 1's silicates are often seen in emission \citep{hat15}.  We find that the silicate feature has a more significant impact on EW(OH119) than the Balmer decrement.  A relationship between EW(OH65) and $S_{\rm{sil}}$ has also been shown in \citet{gon15}.  As discussed in Section \ref{subsec:silicates}, we find a high correlation coefficient ($r$ = 0.60) but a p-value greater than 0.05.  The high p-value in this situation could be partially attributed to the low sample size (eight galaxies).  Therefore, our sample does not contradict the correlation between EW(OH65) and $S_{\rm{sil}}$ found in \citet{gon15}, but is too small to be conclusive.  With only a small sample (four galaxies), this study finds no correlation between EW(OH65) and the Balmer decrement as well.  The results for EW(OH79) oppose the idea that a IR dust tracer is preferable as that doublet shows a stronger correlation with the optical Balmer decrement.

There appears to be some connection between the strength of the CO(16-15) line and how obscured the galaxy is.  The spectral separation between OH163 and the CO(16-15) transition is small enough that CO(16-15) is within the spectral range of the OH163 observations.  CO(16-15) is too weak to be detected in the majority of the sample; however, some examples of galaxies where the transition is clearly visible are NGC 253, NGC 1068, NGC 3079, NGC 4418, and NGC 4945.  The transition is very prominent in NGC 4418, which has a deeply buried Seyfert 2 nucleus and the second deepest silicate absorption in the sample ($S_{\rm{sil}}$ = $-$4.00), while it is not so strong in NGC 253, which is a starburst.  The upper level of CO(16-15) is $\sim$750 K, so a very warm and dense region is needed to produce this emission.  In our sample, CO(16-15) is much more likely to be observed in the most obscured galaxies; however, it is not guaranteed to find the transition in these galaxies.  IRAS 17208-0014 has the strongest silicate absorption ($S_{\rm{sil}}$ = $-$4.07); however, CO(16-15) is not visible in the spectrum.

\section{Summary} \label{sec:summary}

In this paper, we investigate possible relationships between the strength of an AGN/starburst component in the host galaxy with the EW values from a series of six OH doublets (OH65, OH71, OH79, OH84, OH119, OH163) in a sample of 178 galaxies.  The sample contains a wide range of optical spectral classifications $-$ a complete range of Seyfert-types, LINERs, and non-active star-forming galaxies.  The sample was observed using the PACS instrument on the \textit{Herschel Space Observatory} and collected from \textit{Hershel Science Archive} and the optical spectral classifications come from the \citet{ver10} catalog.  

We look for trends within the OH data itself, and with galaxy observables.  To probe the spectral type of the galaxy, brightness of an AGN/starburst component, and radiation field of the galaxy, we use many observables.  These include various IR emission-line ratios, various IR emission-line luminosities, the intrinsic 2-10 keV X-ray luminosity, and the 25$\mu$m/60$\mu$m IRAS flux density ratio as a proxy for dust temperature.  We use the strength of the silicate 9.7$\mu$m feature and the Balmer decrement to probe dust extinction.

The main results of this study are as follows:
\begin{enumerate}
    \item In our sample, we find OH71 to be always in absorption, OH65 and OH84 to be primarily in absorption, OH119 to have a comparable number of both emission and absorption observations, OH79 to be primarily in emission, and OH163 to be always in emission.  
    \item For OH79, OH119, and OH163, we find that galaxies containing a bright AGN creating a warm, dense environment are more likely to drive these three lines into strong emission.  For galaxies that have an obscured AGN or no AGN, these show strong absorption (OH119), weak emission (OH163), and a mixture of weak absorption and weak emission (OH79).  These differences could be explained by their method of excitation: collisional excitation (OH119), radiative pumping (OH163), and a mix of both processes (OH79).  
    \item The results for EW(OH65) suggest that there is a possibility that absorption becomes stronger with increasing AGN strength.  This could mean that a warm ($\sim$300 K), dense environment is required to populate the ground levels to enable these transitions, and a less extreme environment (e.g. star-forming galaxies) may not be able to excite a significant number of OH molecules to generate the OH65 transition.  However, this is only based on samples of five galaxies, so more data are needed to confirm this relationship.
    \item We find a relationship between dust extinction and the OH79 and OH119 doublets.  This relationship is seen with both the silicate feature and Balmer decrement for OH79 and OH119.  EW(OH79) has a better correlation with the Balmer decrement, while EW(OH119) has a stronger relationship with $S_{\rm{sil}}$.  For these doublets, we see emission weaken/absorption strengthen as the dust extinction becomes stronger, suggesting that the warm and dense molecular gas producing the OH emission/absorption is associated with a dusty environment.
    \item No correlations are found between either EW(OH71) or EW(OH84) with any of the AGN, radiation field hardness, dust temperature, and dust extinction tracers.
    \item P-Cygni features are observed in $\sim$10\% of the galaxies in our sample.  We find that one galaxy (Circinus) has a reverse P-Cygni feature, which indicates a molecular inflow.
\end{enumerate}

This work is based on observations made with the \textit{Herschel} and the \textit{Spitzer} IR space observatories, whose observations are at the base of the present study.  Future missions SPICA (e.g. \citealt{gon17b, spi17, roe18}) and NASA Origins (e.g. \citealt{mei18}) will be able to extend this work to even larger samples of galaxies, both in the local and distant Universe.

\acknowledgments

We thank the referee for providing valuable comments that helped improve the paper.  PACS has been developed by a consortium of institutes led by MPE (Germany) and including UVIE (Austria); KU Leuven, CSL, IMEC (Belgium); CEA, LAM (France); MPIA (Germany); INAF-IFSI/OAA/OAP/OAT, LENS, SISSA (Italy); IAC (Spain). This development has been supported by the funding agencies BMVIT (Austria), ESA-PRODEX (Belgium), CEA/CNES (France), DLR (Germany), ASI/INAF (Italy), and CICYT/MCYT (Spain).  This research has made use of the NASA/IPAC Extragalactic Database (NED) which is operated by the Jet Propulsion Laboratory, California Institute of Technology, under contract with the National Aeronautics and Space Administration.  This research made use of Astropy,\footnote{http://www.astropy.org} a community-developed core Python package for Astronomy \citep{ast13, ast18}.  JAFO and LS acknowledge financial support by the Agenzia Spaziale Italiana (ASI) under the research contract 2018-31-HH.0.  MPS acknowledges support from the Comunidad de Madrid through Atracci\'on de Talento Investigador Grant 2018-T1/TIC-11035 and STFC through grants ST/N000919/1 and ST/N002717/1.

\vspace{5mm}
\facilities{\textit{Hershel}/PACS, \textit{Spitzer}/IRS, \textit{IRAS}.}

\software{Astropy \citep{ast13, ast18}, HIPE (v13.0.0; \citealt{ott10}), IPython \citep{per07}, Matplotlib \citep{hun07}, NumPy \citep{oli06, van11}, Scikit-learn \citep{ped11}, SciPy \citep{oli07, mil11, vir20}
                    }

\appendix

\section{Fitting Statistics}

\startlongtable
\begin{deluxetable*}{lcccccc}
\tabletypesize{\scriptsize}
\tablecolumns{7}
\tablecaption{Best-fit Parameters for EW(OH) vs. Emission-Line Ratio Fluxes \label{tab:line_ratio_flux_stats}}
\tablehead{
  \colhead{Parameters} &
  \colhead{OH65} &
  \colhead{OH71} & 
  \colhead{OH79} &
  \colhead{OH84} &
  \colhead{OH119} &
  \colhead{OH163} \\
  \colhead{(1)} & \colhead{(2)} & \colhead{(3)} & \colhead{(4)} & \colhead{(5)} & \colhead{(6)} & \colhead{(7)}}
\startdata
\multicolumn{7}{c}{\large{\textbf{log\bigg($\frac{\rm{[Ne\ V]14.32\mu m}}{\rm{[Ne\ II]12.81\mu m}}$\bigg)}}} \\
\hline\hline
\tiny{slope} & \tiny{ } & \tiny{ } & \tiny{0.017 $\pm$ 0.003} & \tiny{ } & \tiny{0.07 $\pm$ 0.01} & \tiny{0.019 $\pm$ 0.009} \\
\tiny{Y-intercept} & \tiny{ } & \tiny{ } & \tiny{0.020 $\pm$ 0.003} & \tiny{ } & \tiny{0.06 $\pm$ 0.01} & \tiny{0.06 $\pm$ 0.01} \\
\tiny{Correlation Coefficient} & \tiny{ } & \tiny{ } & \tiny{0.50} & \tiny{ } & \tiny{0.52} & \tiny{0.39} \\
\tiny{p-value} & \tiny{ } & \tiny{ } & \tiny{0.007***} & \tiny{ } & \tiny{0.0025***} & \tiny{0.14} \\
\hline
\multicolumn{7}{c}{\large{\textbf{log\bigg($\frac{\rm{[Ne\ V]24.32\mu m}}{\rm{[Ne\ II]12.81\mu m}}$\bigg)}}} \\
\hline\hline
\tiny{slope} & \tiny{ } & \tiny{ } & \tiny{0.006 $\pm$ 0.004} & \tiny{ } & \tiny{0.05 $\pm$ 0.02} & \tiny{0.059 $\pm$ 0.008} \\
\tiny{Y-intercept} & \tiny{ } & \tiny{ } & \tiny{0.013 $\pm$ 0.003} & \tiny{ } & \tiny{0.03 $\pm$ 0.01} & \tiny{0.087 $\pm$ 0.06} \\
\tiny{Correlation Coefficient} & \tiny{ } & \tiny{ } & \tiny{0.18} & \tiny{ } & \tiny{0.10} & \tiny{0.69} \\
\tiny{p-value} & \tiny{ } & \tiny{ } & \tiny{0.43} & \tiny{ } & \tiny{0.61} & \tiny{0.02**} \\
\hline
\multicolumn{7}{c}{\large{\textbf{log\bigg($\frac{\rm{[Ne\ III]15.56\mu m}}{\rm{[Ne\ II]12.81\mu m}}$\bigg)}}} \\
\hline\hline
\tiny{slope} & \tiny{$-$0.016 $\pm$ 0.009} & \tiny{ } & \tiny{0.021 $\pm$ 0.004} & \tiny{0.014 $\pm$ 0.008} & \tiny{0.10 $\pm$ 0.02} & \tiny{0.05 $\pm$ 0.01} \\
\tiny{Y-intercept} & \tiny{$-$0.021 $\pm$ 0.008} & \tiny{ } & \tiny{0.018 $\pm$ 0.003} & \tiny{$-$0.005 $\pm$ 0.006} & \tiny{0.05 $\pm$ 0.01} & \tiny{0.07 $\pm$ 0.01} \\
\tiny{Correlation Coefficient} & \tiny{$-$0.24} & \tiny{ } & \tiny{0.46} & \tiny{0.28} & \tiny{0.61} & \tiny{0.43} \\
\tiny{p-value} & \tiny{0.54} & \tiny{ } & \tiny{0.002*} & \tiny{0.38} & \tiny{9.4e-06***} & \tiny{0.03*} \\
\hline
\multicolumn{7}{c}{\large{\textbf{log\bigg($\frac{\rm{[O\ IV]25.89\mu m}}{\rm{[Ne\ II]12.81\mu m}}$\bigg)}}} \\
\hline\hline
\tiny{slope} & \tiny{ } & \tiny{ } & \tiny{0.004 $\pm$ 0.002} & \tiny{0.012 $\pm$ 0.005} & \tiny{0.07 $\pm$ 0.01} & \tiny{0.026 $\pm$ 0.008} \\
\tiny{Y-intercept} & \tiny{ } & \tiny{ } & \tiny{0.008 $\pm$ 0.002} & \tiny{$-$0.001 $\pm$ 0.006} & \tiny{0.010 $\pm$ 0.008} & \tiny{0.06 $\pm$ 0.01} \\
\tiny{Correlation Coefficient} & \tiny{ } & \tiny{ } & \tiny{0.37} & \tiny{0.07} & \tiny{0.45} & \tiny{0.39} \\
\tiny{p-value} & \tiny{ } & \tiny{ } & \tiny{0.02*} & \tiny{0.87} & \tiny{0.005*} & \tiny{0.08} \\
\hline
\multicolumn{7}{c}{\large{\textbf{log\bigg($\frac{\rm{[O\ IV]25.89\mu m}}{\rm{[O\ III]88.36\mu m}}$\bigg)}}} \\
\hline\hline
\tiny{slope} & \tiny{ } & \tiny{ } & \tiny{0.005 $\pm$ 0.003} & \tiny{ } & \tiny{0.08 $\pm$ 0.02} & \tiny{0.022 $\pm$ 0.007} \\
\tiny{Y-intercept} & \tiny{ } & \tiny{ } & \tiny{0.006 $\pm$ 0.002} & \tiny{ } & \tiny{$-$0.01 $\pm$ 0.01} & \tiny{0.043 $\pm$ 0.007} \\
\tiny{Correlation Coefficient} & \tiny{ } & \tiny{ } & \tiny{0.40} & \tiny{ } & \tiny{0.63} & \tiny{0.36} \\
\tiny{p-value} & \tiny{ } & \tiny{ } & \tiny{0.08} & \tiny{ } & \tiny{0.0006***} & \tiny{0.14} \\
\enddata
\tablecomments{
  Col. (1): Linear fitting parameters.
  Col. (2)$-$(7): Results from the linear fit for each OH transition, and the Spearman rank-order correlation coefficient and the p-value to test for non-correlation.  *: indicates p $<$ 0.05; however, the observable is an uncertain predictor of EW(OH) due to $r$ $<$ 0.5. **: indicates a weak correlation with $r$ $\geq$ 0.5 and 0.01 $<$ p $<$ 0.05. ***: indicates a significant correlation with $r$ $\geq$ 0.5 and p $<$ 0.01.  Blank spaces in the table indicate that there is less than 8 galaxies so no fit was made.}
\end{deluxetable*}

\startlongtable
\begin{deluxetable*}{lcccccc}
\tabletypesize{\scriptsize}
\tablecolumns{7}
\tablecaption{Best-fit Parameters for EW(OH) vs. the X-ray and IR Emission-Line Luminosities \label{tab:luminosity_stats}}
\tablehead{
  \colhead{Parameters} &
  \colhead{OH65} &
  \colhead{OH71} & 
  \colhead{OH79} & 
  \colhead{OH84} &
  \colhead{OH119} &
  \colhead{OH163} \\
  \colhead{(1)} & \colhead{(2)} & \colhead{(3)} & \colhead{(4)} & \colhead{(5)} & \colhead{(6)} & \colhead{(7)}}
\startdata
\multicolumn{7}{c}{\large{\textbf{log($L_{\rm{2-10keV}}$)}}} \\
\hline\hline
\tiny{Slope} & \tiny{$-$0.009 $\pm$ 0.004} & \tiny{ } & \tiny{0.05 $\pm$ 0.002} & \tiny{$-$0.2 $\pm$ 0.6} & \tiny{0.16 $\pm$ 0.08} & \tiny{0.06 $\pm$ 0.03} \\
\tiny{Y-intercept} & \tiny{0.38 $\pm$ 0.18} & \tiny{ } & \tiny{$-$1.88 $\pm$ 0.65} & \tiny{7.89 $\pm$ 25.42} & \tiny{$-$6.70 $\pm$ 3.55} & \tiny{$-$2.48 $\pm$ 1.13} \\
\tiny{Correlation Coefficient} & \tiny{$-$0.39} & \tiny{ } & \tiny{0.41} & \tiny{$-$0.07} & \tiny{0.35} & \tiny{0.36} \\
\tiny{p-value} & \tiny{0.30} & \tiny{ } & \tiny{0.014*} & \tiny{0.83} & \tiny{0.03*} & \tiny{0.09} \\
\hline
\multicolumn{7}{c}{\large{\textbf{log($L_{\rm{[O\ IV]25.89\mu m}}$)}}} \\
\hline\hline
\tiny{Slope} & \tiny{ } & \tiny{ } & \tiny{0.005 $\pm$ 0.001} & \tiny{0.006 $\pm$ 0.005} & \tiny{0.035 $\pm$ 0.007} & \tiny{0.017 $\pm$ 0.004} \\
\tiny{Y-intercept} & \tiny{ } & \tiny{ } & \tiny{$-$0.18 $\pm$ 0.06} & \tiny{$-$0.24 $\pm$ 0.19} & \tiny{$-$1.44 $\pm$ 0.27} & \tiny{$-$0.65 $\pm$ 0.17} \\
\tiny{Correlation Coefficient} & \tiny{ } & \tiny{ } & \tiny{0.31} & \tiny{$-$0.38} & \tiny{0.31} & \tiny{0.56} \\
\tiny{p-value} & \tiny{ } & \tiny{ } & \tiny{0.06} & \tiny{0.35} & \tiny{0.06} & \tiny{0.007***} \\
\hline
\multicolumn{7}{c}{\large{\textbf{log($L_{\rm{[Ne\ V]14.32\mu m}}$)}}} \\
\hline\hline
\tiny{Slope} & \tiny{ } & \tiny{ } & \tiny{0.010 $\pm$ 0.002} & \tiny{ } & \tiny{0.039 $\pm$ 0.007} & \tiny{0.013 $\pm$ 0.005} \\
\tiny{Y-intercept} & \tiny{ } & \tiny{ } & \tiny{$-$0.41 $\pm$ 0.08} & \tiny{ } & \tiny{$-$1.56 $\pm$ 0.27} & \tiny{$-$0.47 $\pm$ 0.22} \\
\tiny{Correlation Coefficient} & \tiny{ } & \tiny{ } & \tiny{0.33} & \tiny{ } & \tiny{0.27} & \tiny{0.35} \\
\tiny{p-value} & \tiny{ } & \tiny{ } & \tiny{0.09} & \tiny{ } & \tiny{0.13} & \tiny{0.19} \\
\hline
\multicolumn{7}{c}{\large{\textbf{log($L_{\rm{[Ne\ V]24.32\mu m}}$)}}} \\
\hline\hline
\tiny{Slope} & \tiny{ } & \tiny{ } & \tiny{0.009 $\pm$ 0.006} & \tiny{ } & \tiny{0.009 $\pm$ 0.005} & \tiny{0.04 $\pm$ 0.02} \\
\tiny{Y-intercept} & \tiny{ } & \tiny{ } & \tiny{$-$0.35 $\pm$ 0.24} & \tiny{ } & \tiny{$-$0.35 $\pm$ 0.19} & \tiny{$-$1.51 $\pm$ 0.82} \\
\tiny{Correlation Coefficient} & \tiny{ } & \tiny{ } & \tiny{$-$0.13} & \tiny{ } & \tiny{$-$0.002} & \tiny{0.17} \\
\tiny{p-value} & \tiny{ } & \tiny{ } & \tiny{0.56} & \tiny{ } & \tiny{0.99} & \tiny{0.61} \\
\enddata
\tablecomments{
  Col. (1): Linear fitting parameters.
  Col. (2)$-$(7): Results from the linear fit for each OH transition, and the Spearman rank-order correlation coefficient and the p-value to test for non-correlation.  *: indicates p $<$ 0.05; however, the observable is an uncertain predictor of EW(OH) due to $r$ $<$ 0.5. **: indicates a weak correlation with $r$ $\geq$ 0.5 and 0.01 $<$ p $<$ 0.05. ***: indicates a significant correlation with $r$ $\geq$ 0.5 and p $<$ 0.01.  Blank spaces in the table indicate that there is less than 8 galaxies so no fit was made.}
\end{deluxetable*}

\startlongtable
\begin{deluxetable*}{lcccc}
\tabletypesize{\scriptsize}
\tablecolumns{3}
\tablecaption{Best-fit Parameters for Figure \ref{fig:iras_plot} (EW(OH) vs. IRAS $f_{25\mu\rm{m}}$/$f_{60\mu\rm{m}}$ Dust Temperature Tracer) \label{tab:flux_stats}}
\tablehead{
  \colhead{OH Doublet} &
  \colhead{Slope} & 
  \colhead{Y-intercept} &
  \colhead{Correlation Coefficient} &
  \colhead{p-value} \\
  \colhead{} & \colhead{(m)} & \colhead{(b)} & \colhead{ } & \colhead{ } \\
  \colhead{(1)} & \colhead{(2)} & \colhead{(3)} & \colhead{(4)} & \colhead{(5)}}
\startdata
OH65 & 0.08 $\pm$ 0.05 & 0.06 $\pm$ 0.06 & 0.11 & 0.76 \\
OH71 &   &   &   &   \\
OH79 & 0.052 $\pm$ 0.006 & 0.045 $\pm$ 0.006 & 0.50 & 0.001*** \\
OH84 & 0.03 $\pm$ 0.01 & 0.02 $\pm$ 0.01 & 0.12 & 0.71 \\
OH119 & 0.22 $\pm$ 0.03 & 0.15 $\pm$ 0.02 & 0.57 & 0.00005*** \\
OH163 & $-$0.004 $\pm$ 0.015 & 0.024 $\pm$ 0.014 & $-$0.03 & 0.89 \\
\enddata
\tablecomments{
  Col. (1): OH doublet.
  Col. (2): Slope of the best-fit line.
  Col. (3): Y-intercept of the best-fit line.
  Col. (4): The Spearman rank-order correlation coefficient.
  Col. (5): The p-value to test for non-correlation.  *: indicates p $<$ 0.05; however, the observable is an uncertain predictor of EW(OH) due to $r$ $<$ 0.5. **: indicates a weak correlation with $r$ $\geq$ 0.5 and 0.01 $<$ p $<$ 0.05. ***: indicates a significant correlation with $r$ $\geq$ 0.5 and p $<$ 0.01.  For columns (2)-(5), blank rows in the table indicate that there is less than 8 galaxies so no fit was made.}
\end{deluxetable*}

\startlongtable
\begin{deluxetable*}{lcccccc}
\tabletypesize{\scriptsize}
\tablecolumns{7}
\tablecaption{Best-fit Parameters for EW(OH) vs. Dust Extinction \label{tab:extinction_stats}}
\tablehead{
  \colhead{Parameters} &
  \colhead{OH65} &
  \colhead{OH71} & 
  \colhead{OH79} &
  \colhead{OH84} &
  \colhead{OH119} &
  \colhead{OH163} \\
  \colhead{(1)} & \colhead{(2)} & \colhead{(3)} & \colhead{(4)} & \colhead{(5)} & \colhead{(6)} & \colhead{(7)}}
\startdata
\multicolumn{7}{c}{\large{\textbf{$S_{\rm{sil}}$}}} \\
\hline\hline
\tiny{Slope} & \tiny{0.005 $\pm$ 0.003} & \tiny{ } & \tiny{0.007 $\pm$ 0.001} & \tiny{0.006 $\pm$ 0.003} & \tiny{0.041 $\pm$ 0.004} & \tiny{0.002 $\pm$ 0.003} \\
\tiny{Y-intercept} & \tiny{$-$0.002 $\pm$ 0.006} & \tiny{ } & \tiny{0.009 $\pm$ 0.002} & \tiny{$-$0.013 $\pm$ 0.005} & \tiny{0.027 $\pm$ 0.006} & \tiny{0.04 $\pm$ 0.01} \\
\tiny{Correlation Coefficient} & \tiny{0.60} & \tiny{ } & \tiny{0.42} & \tiny{0.38} & \tiny{0.59} & \tiny{0.02} \\
\tiny{p-value} & \tiny{0.12} & \tiny{ } & \tiny{0.02*} & \tiny{0.31} & \tiny{0.0001***} & \tiny{0.92} \\
\hline
\multicolumn{7}{c}{\large{\textbf{Balmer Decrement}}} \\
\hline\hline
\tiny{Slope} & \tiny{ } & \tiny{ } & \tiny{$-$0.0011 $\pm$ 0.0003} & \tiny{ } & \tiny{$-$0.010 $\pm$ 0.004} & \tiny{$-$0.0021 $\pm$ 0.0009} \\
\tiny{Y-intercept} & \tiny{ } & \tiny{ } & \tiny{0.018 $\pm$ 0.004} & \tiny{ } & \tiny{0.09 $\pm$ 0.03} & \tiny{0.06 $\pm$ 0.02} \\
\tiny{Correlation Coefficient} & \tiny{ } & \tiny{ } & \tiny{$-$0.63} & \tiny{ } & \tiny{$-$0.47} & \tiny{0.07} \\
\tiny{p-value} & \tiny{ } & \tiny{ } & \tiny{0.002***} & \tiny{ } & \tiny{0.014*} & \tiny{0.80} \\
\hline
\enddata
\tablecomments{
  Col. (1): Linear fitting parameters.
  Col. (2)$-$(7): Results from the linear fit for each OH transition, and the Spearman rank-order correlation coefficient and the p-value to test for non-correlation.  *: indicates p $<$ 0.05; however, the observable is an uncertain predictor of EW(OH) due to $r$ $<$ 0.5. **: indicates a weak correlation with $r$ $\geq$ 0.5 and 0.01 $<$ p $<$ 0.05. ***: indicates a significant correlation with $r$ $\geq$ 0.5 and p $<$ 0.01. Blank spaces in the table indicate that there is less than 8 galaxies so no fit was made.}
\end{deluxetable*}

\begin{deluxetable*}{lccccc}
\tabletypesize{\scriptsize}
\tablecolumns{6}
\tablecaption{Bivariate Regression Results \label{tab:biv_table}}
\tablehead{
  \colhead{OH Doublet} &
  \colhead{log([O~\textsc{IV}] 25.89$\mu$m/[Ne~\textsc{II}] 12.81$\mu$m) Coeff.} & 
  \colhead{log($L_{\rm{2-10keV}}$) Coeff.} & 
  \colhead{$S_{\rm{sil}}$ Coeff.} & 
  \colhead{Y-Intercept} &
  \colhead{$R^2$} \\
  \colhead{(1)} & \colhead{(2)} & \colhead{(3)} & \colhead{(4)}}
\startdata
OH65 &  &  &  &  &  \\
OH71 &  &  &  &  &  \\
OH79 & 0.0026 & $-$0.0007 & 0.0088 & 0.0437 & 0.61 \\
OH84 &  &  &  &  &  \\
OH119 & $-$0.005 & 0.018 & 0.039 & $-$0.755 & 0.91 \\
OH163 & $-$0.005 & 0.003 & 0.005 & $-$0.070 & 0.04 \\
\enddata
\tablecomments{
  Col. (1): OH doublet.
  Col. (2)-(4): Coefficients for the log([O~\textsc{IV}] 25.89$\mu$m/[Ne~\textsc{II}] 12.81$\mu$m), log($L_{\rm{2-10keV}}$), and $S_{\rm{sil}}$ fits.
  Col. (5): Y-intercept for the bivariate fit.
  Col. (4): $R^2$ for the bivariate fit.  Blank spaces in the table indicate that there is less than 8 galaxies so no fit was made.}
\end{deluxetable*}

\end{document}